\documentclass[preprint,aps,prd,amsmath,amssymb,nofootinbib]{revtex4}
\usepackage{graphicx}
\usepackage{color}
\usepackage[latin1]{inputenc}
\usepackage{ulem}
\newcommand{\be}{\begin{eqnarray}}
\newcommand{\beq}{\begin{equation}}
\newcommand{\eeq}{\end{equation}}
\newcommand{\ee}{\end{eqnarray}}

\newcommand{\bmp}{\noindent\begin{minipage}{16cm}}
\newcommand{\emp}{\end{minipage}\vskip 7mm} 

\usepackage{graphicx}
\usepackage{dcolumn}
\usepackage{bm}
\usepackage{amsmath}
\usepackage{amsfonts}
\usepackage{bbm}
\usepackage{subfigure}

\def\drawbox#1#2{\hrule height#2pt
        \hbox{\vrule width#2pt height#1pt \kern#1pt
              \vrule width#2pt}
              \hrule height#2pt}

\def\Asym#1#2{\vcenter{\vbox{\drawbox{#1}{#2}
              \kern-#2pt 
              \drawbox{#1}{#2}}}}

\def\s{{\,\rm s}}
\def\g{{\,\rm g}}
\def\eV{\,{\rm eV}}
\def\keV{\,{\rm keV}}
\def\MeV{\,{\rm MeV}}
\def\GeV{\,{\rm GeV}}

\def\sv{\left<\sigma v\right>}
\def\({\left(}
\def\){\right)}
\def\cm{{\,\rm cm}}
\def\K{{\,\rm K}}

\begin{document}

\title{Strong Interactive Massive Particles from a Strong Coupled Theory}
\author{Maxim Yu. {\sc Khlopov}}\email{maxim.khlopov@roma1.infn.it}
\author{Chris {\sc Kouvaris}}\email{kouvaris@nbi.dk}
 \affiliation{$^*$Center for Cosmoparticle physics ``Cosmion",
125047, Moscow, Russia\\
Moscow Engineering Physics Institute, 115409 Moscow, Russia, and \\
APC laboratory 10, rue Alice Domon et Léonie Duquet 75205 Paris
Cedex 13, France, \\
 $^{\dagger}$CERN Theory Division, CH-1211 Geneva 23,
 Switzerland, \\
University of Southern Denmark, Campusvej 55, DK-5230 Odense,
Denmark and \\
The Niels Bohr Institute, Blegdamsvej 17, DK-2100 Copenhagen,
Denmark}

\begin{abstract}
    {Minimal walking technicolor models can provide a nontrivial solution
for cosmological dark matter, if the lightest technibaryon is
doubly charged. Technibaryon asymmetry generated in the early
Universe is related to baryon asymmetry and it is possible to
create excess of techniparticles with charge ($-2$). These
excessive techniparticles are all captured by $^4He$, creating
\emph{techni-O-helium} $tOHe$ ``atoms'', as soon as $^4He$ is
formed in Big Bang Nucleosynthesis. The interaction of
techni-O-helium with nuclei opens new paths to the creation of
heavy nuclei in Big Bang Nucleosynthesis. Due to the large mass of
technibaryons, the $tOHe$ ``atomic'' gas decouples from the
baryonic matter and plays the role of dark matter in large scale
structure formation, while structures in small scales are
suppressed.
Nuclear interactions with matter slow down cosmic
techni-O-helium in Earth
below the threshold of underground dark matter detectors, thus
escaping severe CDMS constraints. On the other hand, these nuclear
interactions
are not sufficiently strong to
exclude this form of Strongly Interactive Massive Particles by
constraints from the XQC experiment. Experimental tests of this
hypothesis are possible in search for $tOHe$ in balloon-borne
experiments (or on the ground) and for its charged techniparticle
constituents in cosmic rays and accelerators. The $tOHe$ ``atoms''
can cause cold nuclear transformations in matter and might form
anomalous isotopes, offering possible ways to exclude (or prove?)
their existence.}
\end{abstract}


\maketitle
\section{Introduction}
The question of the existence of new quarks and leptons is among
the most important in the modern high energy physics. This
question has an interesting cosmological aspect. If these quarks
and/or charged leptons are stable, they should be present around
us and the reason for their evanescent nature should be found.

Recently, at least three elementary particle frames for heavy
stable charged quarks and leptons were considered: (a) A heavy
quark and heavy neutral lepton (neutrino with mass above half the
$Z$-boson mass) of a fourth generation \cite{N,Legonkov}, which
can avoid experimental constraints \cite{Q,Okun}, and form
composite dark matter species \cite{I,lom,KPS06,Khlopov:2006dk};
(b) A Glashow's ``Sinister'' heavy tera-quark $U$ and
tera-electron $E$, which can form a tower of tera-hadronic and
tera-atomic bound states with ``tera-helium atoms'' $(UUUEE)$
considered as dominant dark matter \cite{Glashow,Fargion:2005xz};
(c) AC-leptons, based on the approach of almost-commutative
geometry \cite{5,book}, that can form evanescent AC-atoms, playing
the role of dark matter \cite{5,FKS,Khlopov:2006uv}.

In all these recent models, the predicted stable charged particles
escape experimental discovery, because they are hidden in elusive
atoms, composing the dark matter of the modern Universe. It offers
a new solution for the physical nature of the cosmological dark
matter. Here we show that such a solution is possible in the
framework of walking technicolor models
\cite{Sannino:2004qp,Hong:2004td,Dietrich:2005jn,Dietrich:2005wk,Gudnason:2006ug,Gudnason:2006yj}
and can be realized without an {\it ad hoc} assumption on charged
particle excess, made in the approaches (a)-(c).

This approach differs from the idea of dark matter composed of
primordial bound systems of superheavy charged particles and
antiparticles, proposed earlier to explain the origin of Ultra
High Energy Cosmic Rays (UHECR) \cite{UHECR}. To survive to the
present time and to be simultaneously the source of UHECR,
superheavy particles should satisfy a set of constraints, which in
particular exclude the possibility that they possess gauge charges
of the standard model.

The particles considered here, participate in the Standard Model
interactions and we show how the problems, related to various dark
matter scenarios with composite atom-like systems, can find an
elegant solution on the base of the minimal walking technicolor
model.

The approaches (b) and (c) try to escape the problems of free
charged dark matter particles \cite{Dimopoulos:1989hk} by hiding
opposite-charged particles in atom-like bound systems, which
interact weakly with baryonic matter. However, in the case of
charge symmetry, when primordial abundances of particles and
antiparticles are equal, annihilation in the early Universe
suppresses their concentration. If this primordial abundance still
permits these particles and antiparticles to be the dominant dark
matter, the explosive nature of such dark matter is ruled out by
constraints on the products of annihilation in the modern Universe
\cite{Q,FKS}. Even in the case of charge asymmetry with primordial
particle excess, when there is no annihilation in the modern
Universe, binding of positive and negative charge particles is
never complete and positively charged heavy species should retain.
Recombining with ordinary electrons, these heavy positive species
give rise to cosmological abundance of anomalous isotopes,
exceeding experimental upper limits. To satisfy these upper
limits, the anomalous isotope abundance on Earth should be
reduced, and the mechanisms for such a reduction are accompanied
by effects of energy release which are strongly constrained, in
particular, by the data from large volume detectors.

These problems of composite dark matter models \cite{Glashow,5}
revealed in \cite{Q,Fargion:2005xz,FKS,I}, can be avoided, if the
excess of only $-2$ charge $A^{--}$
particles is generated in the early Universe. Here we show that in
walking technicolor models, technilepton and technibaryon excess
is related to baryon excess and the excess of $-2$ charged
particles can appear naturally for a reasonable choice of model
parameters. It distinguishes this case from other composite dark
matter models, since in all the previous realizations, starting
from \cite{Glashow}, such an excess was put by hand to saturate
the observed cold dark matter (CDM) density by composite dark
matter.

After it is formed in Big Bang Nucleosynthesis, $^4He$ screens the
$A^{--}$ charged particles in composite $(^4He^{++}A^{--})$
``atoms''. These neutral primordial nuclear interacting objects
saturate the modern dark matter density and play the role of a
nontrivial form of strongly interacting dark matter
\cite{Starkman,McGuire:2001qj}. The active influence of this type
of dark matter on nuclear transformations seems to be incompatible
with the expected dark matter properties. However, it turns out
that the considered scenario is not easily ruled out \cite{FKS,I}
and challenges the experimental search for techni-O-helium and its
charged techniparticle constituents.

The structure of the present paper is as follows. Starting with a
review of possible dark matter candidates offered by the minimal
walking technicolor model, we reveal the possibility for the
lightest techniparticle(s) to have electric charge $\pm 2$
(Section II).
 In Section III we show how the minimal technicolor model
 can provide substantial excess of techniparticles with electric charge
 $-2$.
In Section IV we show how all these $-2$ charge particles can be
captured by $^4He$, after its formation in the Standard Big Bang
Nucleosynthesis (SBBN),
 making neutral techni-O-helium ``atoms" that can account for the modern
dark matter density. Techni-O-helium catalyzes a path for heavy
element formation in SBBN, but we stipulate in Section IV a set of
arguments, by which the considered scenario can avoid immediate
contradiction with observations. Gas of heavy techni-O-helium
``atoms" decouples from the plasma and radiation only at a
temperature about few hundreds eV, so that small scale density
fluctuations are suppressed and gravitational instability in this
gas develops more close to warm dark matter, rather than to cold
dark matter scenario (subsection A of Section V). We further
discuss in Section V the possibility to detect charged
techniparticle components of cosmic rays (subsection B), effects
of techni-O-helium catalyzed processes in Earth (subsection C),
and possibilities of direct searches for techni-O-helium
(subsection D). The problems, signatures, and possible
experimental tests of the techni-O-helium Universe are considered
in Section VI. Details of our calculations are presented in the
Appendices 1 and 2.
\section{Dark Matter from Walking Technicolor}

The minimal walking technicolor model
\cite{Sannino:2004qp,Hong:2004td,Dietrich:2005jn,Dietrich:2005wk,Gudnason:2006ug,Gudnason:2006yj}
has two techniquarks, i.e. up $U$ and down $D$, that transform
under the adjoint representation of an $SU(2)$ technicolor gauge
group. The global symmetry of the model is an $SU(4)$ that breaks
spontaneously to an $SO(4)$. The chiral condensate of the
techniquarks breaks the electroweak symmetry. There are nine
Goldstone bosons emerging from the symmetry breaking. Three of
them are eaten by the $W$ and the $Z$ bosons. The remaining six
Goldstone bosons are $UU$, $UD$, $DD$ and their corresponding
antiparticles. For completeness $UU$ is
$U^{\top}_{\alpha}CU_{\beta}\delta^{\alpha \beta}$, where $C$ is
the charge conjugate matrix and the Greek indices denote
technicolor states. For simplicity in our notation we omit the
contraction of Dirac and technicolor indices. Since the
techniquarks are in the adjoint representation of the $SU(2)$,
there are three technicolor states. The $UD$ and $DD$ have similar
Dirac and technicolor structure. The pions and kaons which are the
Goldstone bosons in QCD carry no baryon number since they are made
of pairs of quark-antiquark. However in our case, the six
Goldstone bosons carry technibaryon number since they are made of
two techniquarks or two anti-techniquarks. This means that if no
processes violate the technibaryon number, the lightest
technibaryon will be stable. The electric charges of $UU$, $UD$,
and $DD$ are given in general by $y+1$, $y$, and $y-1$
respectively, where $y$ is an arbitrary real number. For any real
value of $y$, gauge anomalies are
cancelled~\cite{Gudnason:2006yj}. The model requires in addition
the existence of a fourth family of leptons, i.e. a ``new
neutrino'' $\nu'$ and a ``new electron'' $\zeta$ in order to
cancel the Witten global anomaly. Their electric charges are in
terms of $y$ respectively $(1-3y)/2$ and $(-1-3y)/2$. The
effective theory of this minimal walking technicolor model has
been presented in~\cite{Gudnason:2006ug,Foadi:2007ue}.

There are several possibilities for a dark matter candidate
emerging from this minimal walking technicolor model. For the case
where $y=1$, the $D$ techniquark (and therefore also the $DD$
boson) become electrically neutral. If one assumes that $DD$ is
the lightest technibaryon, then it is absolutely stable, because
there is no way to violate the technibaryon number apart from the
sphalerons that freeze out close to the electroweak scale. This
scenario was studied in
Refs.~\cite{Gudnason:2006ug,Gudnason:2006yj}. It was shown that
$DD$ can provide the full dark matter density if its mass is of
the order of TeV. The exact value of the mass of $DD$ depends on
the temperature where sphalerons freeze out, and on the ratios
$L/B$ and $L'/B$, where $L$ and $L'$ are the lepton number and the
lepton number of the fourth lepton family respectively, and $B$ is
the baryon number. However, this scenario is ruled out by the CDMS
experiment, if $DD$ accounts for 100$\%$ of the dark matter
density. The reason is that since $DD$ has a Spin Independent (SI)
interaction with nuclei, it can scatter coherently in underground
dark matter detectors, raising the elastic cross section. Such a
cross section is already excluded by CDMS, if we accept that the
local dark matter density ranges between $0.2-0.4$ \GeV$/\cm^3$.
However, if $DD$ is a subdominant component, contributing up to
20$\%$ of the total dark matter density, it cannot yet be ruled
out.

 Within the same model and electric charge
assignment, there is another possibility. Since both techniquarks
and technigluons transform under the adjoint representation of the
$SU(2)$ group, it is possible to have bound states between a $D$
and a technigluon $G$. The object $D^{\alpha}G^{\alpha}$ (where
$\alpha$ denotes technicolor states) is techni-colorless. If such
an object has a Majorana mass, then it can account for the whole
dark matter density without being excluded by CDMS, due to the
fact that Majorana particles have no SI interaction with nuclei
and their non-coherent elastic cross section is very low for the
current sensitivity of detectors~\cite{Kouvaris:2007iq}. We should
emphasize that nonzero Majorana mass means that the technibaryon
number is not protected, as in the previous case. For this
scenario to be true, the bound state of $DG$ should be lighter
than $DD$. The lack of tools in order to study the spectrum of the
theory makes hard to decisively conclude if this is true. Lattice
calculations have difficulties to study objects like $DG$ and
evidently perturbation techniques cannot apply. However, in
studies of Super Yang Mills models with supersymmetry softly
broken, it has been argued that a Majorana mass for the gluino
makes the $\lambda G$ ($\lambda$ being the gluino) lighter than
the $\lambda \lambda$~\cite{Evans:1997jy}. If we transfer directly
these results in our case, it would mean that $DG$ is lighter than
$DD$ as long as $D$ has a Majorana mass. Of course this argument
can be only taken as an indication, since the considered walking
technicolor model is not supersymmetric and because $D$ in
principle has also a Dirac mass. On the other hand, if the
Majorana mass is zero, the above argument cannot be applied. In
this case it is more natural to expect that $DD$ (or $UU$ and
$UD$), which is a Goldstone boson, is the lightest technibaryon.
That might imply that $DG$ is unstable.

Finally, if one choose $y=1/3$, $\nu'$ has zero electric charge.
In this case the heavy fourth Majorana neutrino $\nu'$ can play
the role of a dark matter particle. This scenario was explored
first in~\cite{Kainulainen:2006wq} and later
in~\cite{Kouvaris:2007iq}. It was shown that indeed the fourth
heavy neutrino can provide the dark matter density without being
excluded by CDMS~\cite{Akerib:2004fq} or any other experiment.
This scenario allows the possibility for new signatures of weakly
 interacting massive particle annihilation~\cite{Kouvaris:2007ay}.

In this paper we study a case that resembles mostly the first one
 mentioned above, that is $y=1$ and the Goldstone bosons $UU$,
$UD$, and $DD$ have electric charges 2, 1, and 0 respectively. In
addition for $y=1$, the electric charges of $\nu'$ and $\zeta$ are
respectively $-1$ and $-2$. We are interested in the case where
stable particles with $-2$ electric charge have substantial relic
densities and can capture $^4He^{++}$ nuclei to form a neutral
atom. There are three possibilities for this scenario. The first
one is to have a relic density of $\bar{U}\bar{U}$, which has $-2$
charge. For this to be true we should assume that $UU$ is lighter
than $UD$ and $DD$ and no processes (apart from electroweak
sphalerons) violate the technibaryon number. The second one is to
have abundance of $\zeta$ that again has $-2$ charge and the third
case is to have both $\bar{U}\bar{U}$ (or $DD$ or
$\bar{D}\bar{D}$) and $\zeta$. For the first case to be realized,
$UU$ although charged, should be lighter than both $UD$ and $DD$.
This can happen if one assumes that there is an isospin splitting
between $U$ and $D$. This is not hard to imagine since for the
same reason in QCD the charged proton is lighter than the neutral
neutron. Upon making this assumption, $UD$ and $DD$ will decay
through weak interactions to the lightest $UU$. The technibaryon
number is conserved and therefore $UU$ (or $\bar{U}\bar{U}$) is
stable. Similarly in the second case where $\zeta$ is the abundant
$-2$ charge particle, $\zeta$ must be lighter than $\nu'$ and
there should be no mixing between the fourth family of leptons and
the other three of the Standard Model. The $L'$ number is violated
only by sphalerons and therefore after the temperature falls
roughly below the electroweak scale $\Lambda_{EW}$ and the
sphalerons freeze out, $L'$ is conserved, which means that the
lightest particle, that is $\zeta$ in this case, is absolutely
stable. We assume also that technibaryons decay to Standard Model
particles through Extended Technicolor (ETC) interactions and
therefore $TB=0$. Finally in the third case, we examine the
possibility to have both $L'$ and $TB$ conserved after sphalerons
have frozen out. In this case, the dark matter would be composed
of bound atoms $(^4He^{++}\zeta^{--})$ and either $(^4He^{++}(\bar
U \bar U )^{--})$ or neutral $DD$ (or $\bar{D}\bar{D}$).
 We shall examine the three possibilities separately in the next
section.

\section{The excess of the $\bold{-2}$ charged techni-particles in the early Universe}

The calculation of the  excess of the technibaryons with respect
to the one of the baryons
 was pioneered in
Refs.~\cite{Harvey:1990qw,Barr:1990ca,Khlebnikov:1996vj}. In this
paper we calculate the excess of $\bar{U}\bar{U}$ and $\zeta$
along the lines of~\cite{Gudnason:2006yj}. The technicolor and the
Standard Model particles are in thermal equilibrium as long as the
rate of the weak (and color) interactions is larger than the
expansion of the Universe. In addition, the sphalerons allow the
violation of the technibaryon number $TB$, $B$, $L$, and $L'$ as
long as the temperature of the Universe is higher than roughly
$\Lambda_{EW}$. It is possible through the equations of thermal
equilibrium, sphalerons and overall electric neutrality for the
particles of the Universe, to associate the chemical potentials of
the various particles. Following~\cite{Gudnason:2006yj}, we can
write down the $B$, $TB$, $L$ and $L'$ as
 \beq
  B=12 \mu_{uL}+6\mu_W \label{b}
   \eeq
    \beq
     TB=
\frac{2}{3}((\sigma_{UU}+\sigma_{UD}+\sigma_{DD})\mu_{UU}+(\sigma_{UD}+2\sigma_{DD})\mu_W)
\label{tb} \eeq \beq
 L=4\mu+6\mu_W \label{l}
  \eeq
  \beq
L'=4\sigma_{\zeta}\mu_{\nu'_L}+2\sigma_{\zeta}\mu_W,
\label{ll}\eeq where $\mu_{uL}$, $\mu_W$, $\mu_{\nu'}$, $\mu_{UU}$
are respectively the chemical potentials of the left handed up
quark, $W$, $\nu'$, and $UU$. $\mu$ is the sum of the chemical
potentials of the three left handed neutrinos and the
$\sigma_{\alpha}$ parameters denote statistical factors for the
species $\alpha$ defined as \beq
\sigma_{\alpha}=\frac{6}{4\pi^2}\int_0^{\infty}dx x^2
\cosh^{-2}\left(\frac{1}{2}\sqrt{x^2+(\frac{m_{\alpha}}{T^*})^2}\right)
\quad \text{for fermions}, \label{fermion}\eeq \beq
\sigma_{\alpha}=\frac{6}{4\pi^2}\int_0^{\infty}dx x^2
\sinh^{-2}\left(\frac{1}{2}\sqrt{x^2+(\frac{m_{\alpha}}{T^*})^2}\right)
\quad \text{for bosons}, \label{boson}\eeq where $m_{\alpha}$ is
the mass of $\alpha$ and $T^*$ is the freeze out temperature for
the sphaleron. In the derivation we have assumed for simplicity
that the mass of $\nu'$ and $\zeta$ are very close, so
$\sigma_{\nu'}\approx \sigma_{\zeta}$, and that the Standard Model
particles are massless at $T> \Lambda_{EW}$. The sphaleron
processes and the condition of the overall electric neutrality
impose two extra conditions on the chemical
potentials~\cite{Gudnason:2006yj} \beq
9\mu_{uL}+\frac{3}{2}\mu_{UU}+\mu+\mu_{\nu'}+8\mu_W =0,
\label{spha}\eeq \beq
Q=6\mu_{uL}+(2\sigma_{UU}+\sigma_{UD})\mu_{UU}-2\mu -
6\sigma_{\zeta}\mu_{\nu'}+(\sigma_{UD}-4\sigma_{\zeta}-18)\mu_W+(14+\sigma_{\zeta})\mu_0=0,
\label{q} \eeq where $\mu_0$ is the chemical potential of the
Higgs boson.
 Using Eqs.~(\ref{b}), (\ref{tb}), (\ref{l}),
(\ref{ll}), (\ref{spha}), and (\ref{q}), we can write the ratio of
$TB/B$ as a function of the the ratios $L/B$, $L'/B$ and
statistical factors as \beq
\frac{TB}{B}=-\sigma_{UU}\left(\frac{L'}{B}\frac{1}{3\sigma_{\zeta}}+1+\frac{L}{3B}\right).
\label{tbb}\eeq For the derivation of the above ratio, we have
assumed that the electroweak phase transition is of second order,
which means that the sphalerons processes freeze out at a
temperature slightly lower than the electroweak phase transition.
For this reason we have taken $\mu_0=0$, since the chemical
potential of the Higgs boson in the broken phase should be zero.
The calculation in the case of first order phase transition is
slightly different, but the results are very similar to the ones
of the second order~{\cite{Gudnason:2006yj}}. Furthermore, we have
assumed that the mass differences among $UU$, $UD$, and $DD$ are
not large and therefore we have made the approximation
$\sigma_{DD}\simeq \sigma_{UD}\simeq \sigma_{UU}$. In principle we
do not have to make this approximation. The ratio (\ref{tbb})
would look a bit more complicated but it would not change our
physical conclusions. The mass differences among $UU$, $UD$, and
$DD$ depend on the isospin splitting of the two techniquarks $U$
and $D$. However, if the splitting is not large, as in the case
between up and down quarks in QCD, the mass differences among
$UU$, $UD$, and $DD$ are small compared to the electroweak energy
scale and consequently our approximation is justified. We should
emphasize two points regarding the ratio~(\ref{tbb}). The minus
sign in the right hand side denotes the fact that if the quantity
inside the parenthesis is positive, there is an abundance of
anti-technibaryons and not technibaryons. In the first case that
we investigate, where $UU$ is lighter than $UD$ and $DD$, an
abundance of $\bar{U}\bar{U}$ will provide the charge $-2$
particles that capture the positively charged nucleus of helium in
order to form the neutral dark matter atom. The second point is
that (\ref{tbb}) seems to diverge if we take the limit where the
mass of the $\zeta$ and $\nu'$ becomes infinite. In that case
$\sigma_{\zeta}\rightarrow 0$ and the ratio diverges. However,
$L'$ as seen in (\ref{ll}) depends linearly on $\sigma_{\zeta}$
and therefore in the limit where the mass of the $\zeta$ is very
large $L' \rightarrow 0$, unless $\mu_{\nu'}\rightarrow \infty$,
which is something unnatural. As we mentioned already there are
three different cases that we investigate separately regarding the
production of dark matter. The first case is when $\bar{U}\bar{U}$
is the $-2$ charged particle that will bind with helium to form
neutral atom.
\subsection{The case of $\bold{\bar{U}\bar{U}}$}
\begin{figure}[!tbp]
  \begin{center}
    \mbox{
      \subfigure{\resizebox{!}{4.8cm}{\includegraphics{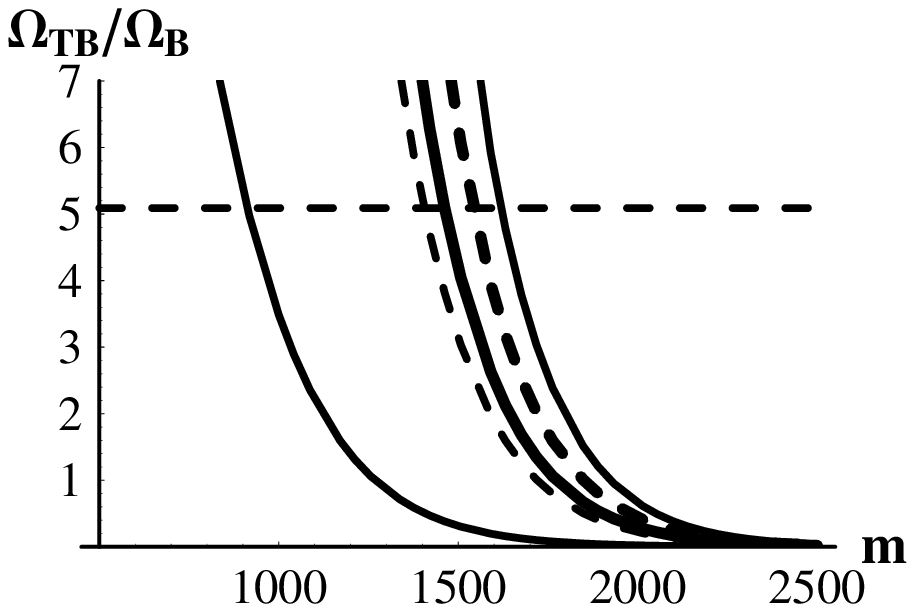}}} \quad
      \subfigure{\resizebox{!}{4.8cm}{\includegraphics{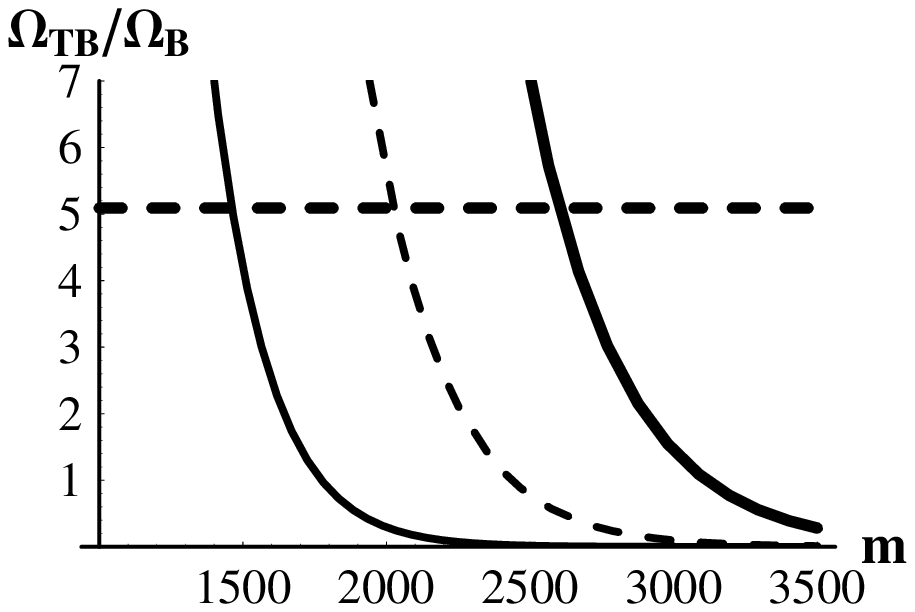}}}
      }
    \caption{\emph{Left Panel}: The ratio of $\Omega_{TB}/\Omega_B$ as
    a function of the mass $m$ (in GeV) of the $\bar{U}\bar{U}$ for several
    values of $\xi$. The first from the left thin solid line
    corresponds to $\xi=0.1$, the thin dashed line to $\xi=1$,
    the thick solid line to $\xi=4/3$, the thick dashed line to
    $\xi=2$ and the last solid line to the right corresponds to
    $\xi=3$. The horizontal dashed line gives the proper ratio of
    dark matter over baryon matter, which is approximately 5. For
    all the curves we have set $T^*=150$ GeV.
     \emph{Right Panel}:The same ratio for fixed $\xi=4/3$, for three different values of the freeze
     out temperature for the sphalerons $T^*$, i.e. $T^*=150$ GeV (thin solid line),
     $T^*=200$ GeV (dashed line), and $T^*=250$ GeV (thick solid line).}
    \label{fig:uu}
    \end{center}
\end{figure}

In this case, $\bar{U}\bar{U}$ is the source of $-2$ charge
particles. The ratio of dark matter produced by the neutral bound
state of $(^4He^{++}(\bar U \bar U )^{--})$ over the baryon matter
is \beq
\frac{\Omega_{DM}}{\Omega_B}=\frac{3}{2}\frac{TB}{B}\frac{m_o}{m_p},
\label{omega1} \eeq where $m_o$ is the mass of the ``dark matter
atom'', which is approximately the mass of $\bar{U}\bar{U}$ plus 4
GeV (the mass of helium) and $m_p$ is the mass of the proton. In
Fig.~(\ref{fig:uu}) we show the ratio $\Omega_{TB}/\Omega_B$ as a
function of the mass $m$ of the $\bar{U}\bar{U}$ for several
values of the parameter $\xi$ and $T^*$ (the sphaleron freeze out
temperature). The parameter $\xi$ is defined as \beq
\xi=\frac{L'}{3B\sigma_{\zeta}}+1+\frac{L}{3B}. \eeq We should
emphasize here that there are two options regarding the new
leptons $\nu'$ and $\zeta$. If $\zeta$ is the lightest between the
two and below the electroweak scale no processes violate $L'$,
then $\zeta$ will contribute to the relic density of $-2$ charge
particles that bind with helium. We study this case later in
subsection C. If $\nu'$ is lighter than $\zeta$, nonzero $L'$
could create problems to our model because we have relic density
of charged $-1$ particles. Therefore, we are forced to assume that
$L'=0$. This is a plausible assumption if one allows mixing
between $\nu'$ and Standard Model leptons, because in that case
$\nu'$ will decay to lighter leptons and $L'=0$. Upon making this
assumption, i.e. $L'=0$, $\xi=1+L/(3B)$. In the left panel of
Fig.~(\ref{fig:uu}) we have chosen $T^*=150$ GeV and several
values of $\xi$ ranging from 0.1 to 3. In principle $\xi$ can take
any positive real value. However, it is logical to assume that the
ratio $L/B$ should be around unity. In fact leptogenesis scenarios
support a ratio of $L/B=1$. However, since currently it is not
possible to know from observations what is the relic density of
the light neutrinos (or antineutrinos), $L/B$ can be also
negative. As it can be seen from the figure, the smaller the value
of $\xi$, the lighter becomes the desired $\bar{U}\bar{U}$ that
can give the dark matter density. For $\xi=0$, the
excess of technibaryons becomes zero within our approximation. We
have plotted several values of $\xi$, namely $\xi=0.1$, 1, 4/3, 2
and 3. In particular, the value $\xi=4/3$ corresponds to $L=B$. We
should stress at this point that for our model to work, $\xi$
should be positive. A negative value of $\xi$ would mean that
there is
an excess for $UU$ (and not $\bar{U}\bar{U}$), which is positively
charged and
 being bound with ordinary electrons plays a role of an anomalous
helium isotope, severely restricted in experimental searches.

\subsection{The case of $\bold{\zeta}$}

Eq.~(\ref{tbb}) gives the ratio of $TB/B$ as a function of $L/B$
and $L'/B$. It can be trivially inverted as \beq
\frac{L'}{B}=-\sigma_{\zeta}\left(\frac{3TB}{\sigma_{UU}B}+3+\frac{L}{B}\right).
\label{tbb2} \eeq In this subsection we investigate the case where
$\zeta$ is the source of $-2$ charge particles that can be
captured by $^4He$. For this to be true, $\zeta$ must be lighter
than $\nu'$ and after sphalerons have frozen out, no other
processes should violate $L'$. The term inside the parenthesis of
(\ref{tbb2}) should be negative in order to have abundance of
$\zeta$ and not anti-$\zeta$. This probably means that a negative
ratio $L/B$ is needed. As we mentioned earlier, for our model to
be realized, only abundance of $-2$ charged or neutral particles
is accepted. Abundance of charged particles with different charges
would cause a serious problem \cite{I,lom,KPS06,Khlopov:2006dk,Fargion:2005xz,FKS}.
We study the case where we have
abundance for both technibaryons and $\zeta$ in the next
subsection. Here, we look at the case where $TB=0$. This can be
realized if below the electroweak scale, ETC processes that
violate $TB$ exist. In such a case, the lightest technibaryon will
decay to lighter Standard Model particles and as a result $TB=0$.
If $TB=0$, the dark matter density that $\zeta$ can provide is
given by \beq
\frac{\Omega_{L'}}{\Omega_B}=\frac{L'}{B}\frac{m_o}{m_p},
\label{omega2} \eeq where $m_o$ in this case is the mass of $\zeta$
plus the 4 GeV mass of $^4He$. Fig.~(\ref{fig:uu}) shows also the
dark matter that $\zeta$ with mass $m$ can provide, if we take
instead of $\xi=1+L/(3B)$, $\xi=-2-2L/(3B)$. For example the curve
with $\xi=4/3$, that in the previous case corresponded to $L=B$,
now corresponds to $L/B=-5$. This identification is possible
because although $\sigma_{\zeta}(m/T^*)$ and $\sigma_{UU}(m/T^*)$
are defined through Eqs.~(\ref{fermion})~and~({\ref{boson})
respectively, for $m>500$ GeV (with $T^*=150$ GeV), the two
parameters are approximately equal.

\subsection{The case of $\bold{\zeta}$ plus $\bold{\bar{U}\bar{U}}$ or $\bold{DD}$ or $\bold{\bar{D}\bar{D}}$}
\begin{figure}[!tbp]
  \begin{center}
    \mbox{
      \subfigure{\resizebox{!}{4.8cm}{\includegraphics{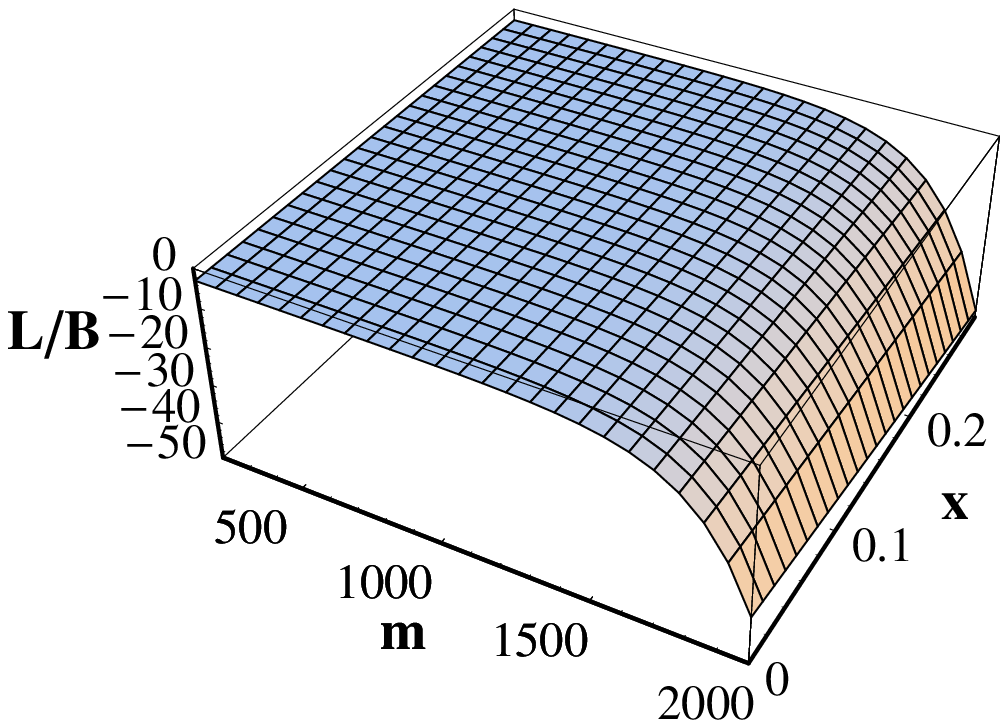}}} \quad
      \subfigure{\resizebox{!}{4.8cm}{\includegraphics{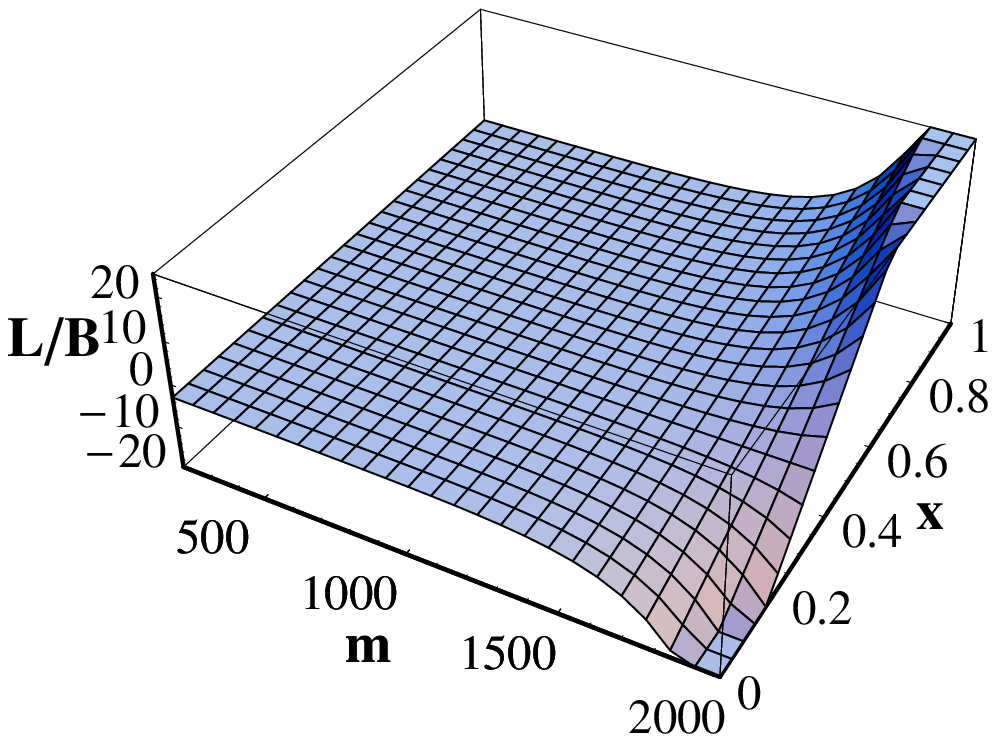}}}
      }
    \caption{\emph{Left Panel}: The value of $L/B$ in terms of the
    mass of the $\zeta$ and the fraction of $DD$ in the dark
    matter density $x$, in order the composition of dark matter to be $x$
    $DD$ and ($1-x$) of $(^4He^{++}\zeta^{--})$.
     \emph{Right Panel}:Same as in the left panel, for the case where the abundant technibaryon is
     either $\bar{D}\bar{D}$ or $\bar{U}\bar{U}$. In this case dark matter is made
     of
      ($1-x$) $(^4He^{++}\zeta^{--})$ and $x$
     $\bar{D}\bar{D}$ or $(^4He^{++}(\bar U \bar U )^{--})$.}
    \label{fig3d}
    \end{center}
\end{figure}
The last case we investigate is the one where below the $T^*$
temperature no processes violate $TB$ and $L'$. This means that
the lightest technibaryon and the lightest fourth family lepton
are stable objects. In particular, we assume that $\zeta$ is
lighter than $\nu'$. As for the technibaryons there are two
options. The first one is to have $DD$ as the lightest
technibaryon. In this case, dark matter will be composed of a
mixture of neutral $DD$ or $\bar{D}\bar{D}$ and bound
$(^4He^{++}\zeta^{--})$ atoms. The second option is to have $UU$
as the lightest technibaryon. In this case the only acceptable
scenario for dark matter is to have a mixture of bound atoms of
$(^4He^{++}\zeta^{--})$ and $(^4He^{++}(\bar U \bar U )^{--})$.
Any other combination, for example an abundance of $UU$ instead of
$\bar{U}\bar{U}$ would create problems in our cosmological model
since $+2$ charged $UU$ represents a form of anomalous helium.
Eq.~(\ref{tbb2}) relates $TB/B$ to $L/B$ and $L'/B$. If both the
technibaryon and $\zeta$ contribute to dark matter, the ratio of
dark matter over baryon matter is given by \beq
\frac{\Omega_{DM}}{\Omega_B}=\frac{\Omega_{TB}}{\Omega_B}+\frac{\Omega_{L'}}{\Omega_B}=5.09.\label{OO}\eeq
The contribution of technibaryon is \beq
\frac{\Omega_{TB}}{\Omega_{B}}=\frac{3}{2}\frac{|TB|}{B}\frac{m_{TB}}{m_p},
\eeq where $m_{TB}$ is the mass of the lightest technibaryon (plus
4 GeV). We have taken the absolute value of $TB$ because if we
have abundance of $DD$, $TB$ is positive, but for $\bar{U}\bar{U}$
or $\bar{D}\bar{D}$, $TB$ is negative. Similarly by
using~(\ref{tbb2}), the contribution of $\zeta$ is \beq
\frac{\Omega_{L'}}{\Omega_{B}}=\sigma_{\zeta}\left|\left(3+\frac{L}{B}+\frac{3TB}{\sigma_{TB}B}\right)\right|
\frac{m_{\zeta}}{m_p},\eeq where $m_{\zeta}$ is the mass of the
$\zeta$ (plus 4 GeV). The term inside the absolute brackets should
be negative, because we want to have abundance of $\zeta$ and not
anti-$\zeta$. Therefore we have to impose the condition \beq
3+\frac{L}{B}+\frac{3TB}{\sigma_{TB}B} <0. \eeq
Using the above equations we can rewrite Eq.~(\ref{OO}) in a more
convenient form \beq
\frac{\Omega_{DM}}{\Omega_B}=\left|\sigma_{\zeta}\left(3+\frac{L}{B}\right)\frac{m_{\zeta}}{m_p}
\pm
\frac{2\sigma_{\zeta}}{\sigma_{TB}}\frac{m_{\zeta}}{m_{TB}}5.09x\right|+5.09x=5.09,
\label{x_definition} \eeq where $x$ denotes the fraction of dark
matter given by the techibaryon. Again, the term inside the
absolute brackets should be negative. If the
excessive technibaryon is the $DD$, the above equation should be
taken by choosing the plus sign for the term that has $\pm$. If
$\bar{U}\bar{U}$ or $\bar{D}\bar{D}$ is the
excessive particle, then the term should be taken with the minus
sign.

We first investigate the case of $DD$ mixing. If $DD$ accounts for
a component of $x100\%$ of the dark matter density, we can express
the ratio $L/B$ as a function of $x$ and the masses of $\zeta$ and
$DD$. In particular, \beq
\frac{L}{B}=-\frac{5.09m_p}{m_{\zeta}\sigma_{\zeta}}[(2z-1)x+1]-3,
\label{DDzeta} \eeq where
$z=(\sigma_{\zeta}m_{\zeta})/(\sigma_{TB} m_{TB})$. If $m_{\zeta}$
is close to $m_{TB}$, then $z=1$ (provided that the masses are
larger than 350 GeV, if $T^*=150$ GeV). The study of $DD$ as a
dark matter candidate in~\cite{Gudnason:2006yj} revealed that $DD$
cannot account for 100$\%$ of dark matter density. If we accept
that the local dark matter density in the vicinity of the earth is
between $0.2-0.4$ \GeV$/\cm^3$, then with the current exposure of
the detectors in CDMS, $DD$ has been ruled out if composes 100$\%$
of the dark matter due to its large cross section. However,
depending on the mass of $DD$ and the local dark matter density,
$DD$ can be a component of dark matter up to $20-30\%$ without
being excluded by CDMS. This means that in Eq.~(\ref{DDzeta}), $x$
should be $0<x<0.3$. In Fig.~(\ref{fig3d}), we plot the value of
$L/B$ as a function of $m_{\zeta}$ (for $z=1$) and $x$ (the
fraction of dark matter provided by $DD$), in order the dark
matter to be a mixture of $x$ $DD$ and $1-x$ of
$(^4He^{++}\zeta^{--})$. There are three points we would like to
stress here. First, it is obvious that the limit $x=0$ corresponds
to the case we studied in the previous subsection. Second, as it
can be seen from Fig.~(\ref{fig3d}), $L/B$ gets a large negative
value very fast as $m_{\zeta}$ increases beyond $1-1.5$ TeV. This
is something probably unnatural since it is expected that $L/B$
should be of order unity. Third, we see that as $x$ increases,
$L/B$ becomes more negative for constant $m_{\zeta}$. This is
because $TB$ is positive and therefore $L/B$ is forced to be more
negative in order to get a relic density of $\zeta$. We should
also emphasize that if $DD$ is as low as 200 GeV, then the CDMS
constraint becomes tougher. Taking the strict constraint, for low
mass for $DD$, $x$ should be less than 0.1 or even 0.05.

We turn now to the case where $TB$ is negative. This means that
the
excessive technibaryon is either $\bar{D}\bar{D}$ or
$\bar{U}\bar{U}$ according to which one is lighter. We can express
as before $L/B$ as \beq
\frac{L}{B}=\frac{5.09m_p}{m_{\zeta}\sigma_{\zeta}}[(2z+1)x-1]-3
\label{UUzeta}. \eeq In the right panel of Fig.~(\ref{fig3d}) we
show again the projected value of $L/B$ in terms of the mass
$m_{\zeta}$ and the fraction $x$ of $\bar{D}\bar{D}$ or
$\bar{U}\bar{U}$. We plot $x$ from 0 to 1. The CDMS constraints
apply for the $\bar{D}\bar{D}$, but not for $\bar{U}\bar{U}$. This
means that if the technibaryon is $\bar{D}\bar{D}$, $x$ should be
at most 0.3. From the figure we see that in order $L/B$ to be of
order unity, a mass for $\zeta$ and $\bar{U}\bar{U}$ between $1-2$
TeV is needed. Again, the limits $x=0$ and $x=1$ were studied in
the previous subsections and correspond to the cases of having
purely $(^4He^{++}\zeta^{--})$ or purely $(^4He^{++}(\bar U \bar
U))^{--}$ respectively.

 The case where both $UU$ and $\zeta$ are stable techniparticles,
offers another possible dark matter scenario. Positively charged
$UU^{++}$ and negatively charged $\zeta^{--}$ can form ``atoms"
$(\zeta^{--} UU^{++})$, which behave as cold dark matter species.
It resembles the $A^{--}C^{++}$ dark matter atoms of the AC-model
\cite{FKS,Khlopov:2006uv}. However, as it took place in the
AC-cosmology, the existence of $+2$ charge species, which remain
free after all the stages of their binding with $-2$ charge
species and $^4He$, has a potential danger of anomalous $He$
overproduction. The solution found in the AC-model
\cite{FKS,Khlopov:2006uv} for this problem of anomalous $He$
involves an additional strict $U(1)$ gauge symmetry, acting on the
AC leptons, and a new Coulomb-like long range interaction between
the AC-leptons, mediated by the corresponding massless $U(1)$
gauge boson. It could be hardly applied to the case of walking
technicolor models. One can have another way to solve this
problem. The abundance of free $+2$ charge techniparticles is
suppressed, if all of them are bound with $-2$ charge
techniparticles. Such a complete binding takes place naurally, if
the excess of $-2$ charge techniparticles is larger, than the one
of $+2$ charge techniparticles. Under these conditions, virtually
all $+2$ charge techniparticles are bound in atoms with $-2$
charge techniparticles, while the residual $-2$ charge
techniparticles bind with $^4He$ in techni-O-helium. The
realization of this scenario in the framework of a walking
technicolor model and the nontrivial cosmological scenarios
involving both cold $(\zeta^{--} UU^{++})$ and warm $(^4He
\zeta^{--})$ forms of composite dark matter goes beyond the scope
of the present paper and it is the subject of a separate work.

Due to strong technicolor interactions, the $TB$ excess of
$\bar{U}\bar{U}$ suppresses strongly the primordial abundance of
the positively charged $UU$ \cite{Chivukula}. However in the case
of weakly interacting $\zeta$ (in the full analogy with the cases
of tera-leptons \cite{Fargion:2005xz} or AC-leptons \cite{FKS}),
its excess does not guarantee a suppression of the corresponding
positively charged antiparticles $\bar{\zeta}$. The stages of
cosmological evolution resulting in virtually complete elimination
of the primordial $\bar{\zeta}$ are stipulated in Appendix 2.

\section{\label{catEHe} The capture of the charged techniparticles by $\bold{^4He}$}

In the Big Bang nucleosynthesis, $^4He$ is formed with an
abundance $r_{He} = 0.1 r_B = 8 \cdot 10^{-12}$ and, being in
excess, binds all the negatively charged techni-species into
atom-like systems. Since the electric charge of $\bar{U}\bar{U}$
(and $\zeta$) is $-2$, neutral ``atoms" are formed, and
$(^4He^{++} \zeta^{--})$ ``atoms" catalyze effectively the $(\zeta
\bar{\zeta})$ binding and annihilation. It turns out \cite{FKS},
that the electromagnetic cascades from this annihilation cannot
influence the light element abundance and the energy release of
this annihilation takes place so early that it does not distort
the CMB spectrum.

At a temperature $T<I_{o} = Z_{TC}^2 Z_{He}^2 \alpha^2 m_{He}/2
\approx 1.6 \MeV,$ where $\alpha$ is the fine structure constant,
and $Z_{TC}=-2$ stands for the electric charge of $\bar{U}\bar{U}$
and/or of $\zeta$, the reaction \beq
\zeta^{--}+^4He^{++}\rightarrow \gamma +(^4He\zeta) \label{EHeIg}
\eeq and/or \beq (\bar{U}\bar{U})^{--}+^4He^{++}\rightarrow \gamma
+(^4He(\bar{U}\bar{U})) \label{EHeIgU} \eeq can take place. In
these reactions neutral techni-O-helium ``atoms" are produced. The
size of these ``atoms" is \cite{I,FKS} \beq R_{o} \sim 1/(Z_{TC}
Z_{He}\alpha m_{He}) \approx 2 \cdot 10^{-13} \cm \label{REHe}
\eeq and it can play a nontrivial catalyzing role in nuclear
transformations. This aspect needs special thorough study, but
some arguments, which we present below following \cite{FKS},
suggest that there should not be contradicting influence on the
primordial element abundance.

For our problem another aspect is also important.  The reactions
~(\ref{EHeIg}), and~(\ref{EHeIgU}) can start only after $^4He$ is
formed, which happens at $T<100 \keV$. Then, inverse reactions of
ionization by thermal photons support Saha-type relationships
between the abundances of these ``atoms", free $-2$ charged
particles, $^4He$ and $\gamma$: \beq \frac{n_{He} n_A}{n_{\gamma}
n_{(HeA)}} = \exp{(-\frac{I_{o}}{T})}. \label{recHeSaha1} \eeq
From now on, by $A^{--}$ we shall denote $\bar{U}\bar{U}$ and
$\zeta$, if the result is independent of their technibaryon or
technilepton nature. When $T$ falls below $T_{rHe} \sim
I_{o}/\ln{\left(n_{\gamma}/n_{He}\right)} \approx I_{o}/27 \approx
60 \keV$, free $A^{--}$ become bound with helium in the reactions
of Eqs.~(\ref{EHeIg}), and~(\ref{EHeIgU}). The fraction of free
$A^{--}$, which forms neutral $(^4He^{++}A^{--})$ depends on the
ratio of the abundance of $A^{--}$ over the one of $^4He$. For
$m_{\zeta}>50 \GeV$ and $m_{TB}>50 \GeV$, this ratio is less than
1. Therefore due to $^4He$ excess, all the $A^{--}$ form
$(^4He^{++}A^{--})$ ``atoms" through the reactions of
Eqs.~(\ref{EHeIg}), and~(\ref{EHeIgU}). Because of this, no free
$A^{--}$ are left at the time when $T \sim$ few keV, where
$(p^{+}A^{--})^-$ ``ions" or $(p^+p^{+}A^{--})$ ``atoms" could
form.

As soon as techni-O-helium $(^4He^{++}\zeta^{--})$ is formed,
heavy antiparticles $\bar{\zeta}$ can penetrate it, expelling
$^4He$ and forming $\zeta$-positronium states $(\zeta
\bar{\zeta})$, in which the antiparticles annihilate. Therefore
the antiparticle $\bar{\zeta}$ can annihilate through formation of
positronium, such as \beq (He\zeta)+\bar{\zeta} \rightarrow (\zeta
\bar{\zeta} \quad \text{annihilation products} ) +^4He.
\label{EHeP} \eeq

\subsection{\label{Hetrap} $\bold{^4He}$ capture of free negative charges}

At a temperature $T \le T_{rHe}$, when the reactions of
(\ref{EHeIg}), and~(\ref{EHeIgU}) dominate, the decrease of the
free $A^{--}$ abundance due to formation of $(^4He^{++}A^{--})$ is
governed by the equation \cite{FKS} \beq
\frac{dr_{A}}{dx}=f_{1He}\left<\sigma v\right> r_{A}r_{He},
\label{Esrecomb} \eeq where $x=T/I_{o}$, $r_{He} = 8 \cdot
10^{-12}$, $\sv$ is given by
$$\sv=  (\frac{4\pi}{3^{3/2}}) \cdot \frac{\bar \alpha^2}{I_{o}\cdot m_{He}}\frac{1}{x^{1/2}},$$
 where $\bar \alpha = Z_{TC} Z_{He} \alpha$, and (see appendix)
$$f_{1He} \approx m_{Pl}I_{o}.$$
The solution of Eq.~(\ref{Esrecomb}) is given by \cite{FKS}
$$r_{A}=r_{A0} \exp\left(-r_{He} J_{He}\right)=
r_{A0}\exp\left(-1.28 \cdot 10^{4} \right),$$ where
$$J_{He}=\int_0^{x_{fHe}} f_{1He}\left<\sigma v\right>dx =$$
\beq = m_{Pl} (\frac{4 \pi}{3^{3/2}}) \cdot \frac{Z_{TC}^2
Z_{He}^2 \alpha^2}{m_{He}} \cdot 2 \cdot \sqrt{x_{fHe}}\approx 1.6
\cdot 10^{15} \label{JHe}. \eeq $x_{fHe}=1/27$ is the value of
$T/I_o$ where the equilibrium abundance of free charged
techniparticles is frozen out and the photodestruction of their
atom like bound states with $He$ does not prevent recombination,
and $m_{Pl}$ is the Planck mass. Thus, virtually all the free
$A^{--}$ are trapped by helium and their remaining abundance
becomes exponentially small.

For particles $Q^-$ with charge $-1$, as for tera-electrons in the
sinister model \cite{Glashow}, $^4He$ trapping results in the
formation of  a positively charged ion $(^4He^{++} Q^-)^+$, result
in dramatic over-production of anomalous hydrogen
\cite{Fargion:2005xz}. Therefore, only the choice of $- 2$
electric charge for stable techniparticles makes it possible to
avoid this problem. In this case, $^4He$ trapping leads to the
formation of neutral {\it OLe-helium} ``atoms"
$(^4He^{++}A^{--})$, which can catalyze the complete elimination
of primordial positively charged anti-technileptons.

\subsection{\label{ANcat} Complete elimination of antiparticles by techni-O-helium catalysis}
For large $m_{\zeta}$\footnote{From now on $m_{\zeta}$ represents
the mass of $\zeta$ and not the mass of $\zeta$ plus 4 GeV.}, the
primordial abundance of antiparticles $\bar{\zeta}$ is not
suppressed. The presence of techni-O-helium $(^4He^{++}
\zeta^{--})$ in this case accelerates the annihilation of these
antiparticles through the formation of $\zeta$-positronium.
Similar to the case of tera-particles considered in
\cite{Fargion:2005xz} and AC-leptons considered in \cite{FKS}, it
can be shown that the products of annihilation cannot cause a
back-reaction, ionizing techni-O-helium and suppressing the
catalysis.

Indeed, energetic particles, created in $(\zeta \bar{\zeta})$
annihilation, interact with the cosmological plasma. In the
development of the  electromagnetic cascade, the creation of
electron-positron pairs in the reaction $\gamma + \gamma
\rightarrow e^+ + e^-$ plays an important role in astrophysical
conditions (see \cite{BL,AV,bookKh} for a review). The threshold
of this reaction puts an upper limit on the energy of the
nonequilibrium photon spectrum in the cascade \beq E_{max} =
a\frac{m_{e}^2}{25T}, \label{Emax} \eeq where the factor $a =
\ln{(15 \Omega_B + 1)} \approx 0.5$ \cite{FKS}.

At a temperature $T>T_{rbHe} = a m^2_e/(25 I_{o}) \approx 1 \keV$,
in the spectrum of the electromagnetic cascade from $(\zeta
\bar{\zeta})$ annihilation, the maximal energy $E_{max} <I_{o}$
and the annihilation products cannot ionize $(^4He^{++} A^{--})$.
So, there is no back reaction of the $(\zeta \bar{\zeta})$
annihilation until $T \sim T_{rbHe}$. At that time, practically
all free $\zeta$ and $\bar{U}\bar{U}$ are bound into
$(^4He^{++}A^{--})$ atoms. For the same reason, electromagnetic
showers induced by annihilation products, having a maximal  energy
below the binding energies of the SBBN nuclei, cannot initiate
reactions of non-equilibrium nucleosynthesis and influence the
abundance of light elements.

In the absence of back-reaction of annihilation products, nothing
prevents the complete elimination of antiparticles $\bar
\zeta^{++}$ by techni-O-helium catalysis. The $\bar \zeta^{++}$
with primordial abundance $r_{\bar{\zeta}}$, can be captured by
techni-O-helium $(^4He^{++} \zeta^{--})$ with abundance $r_{\zeta
He} = r_{\zeta} =
 r_{\bar \zeta} + \kappa$ (here the
technilepton excess $\kappa = 4 \cdot 10^{-12}f_{\zeta}/S_2$ is
given by Eq.~(\ref{sexcess}) of Appendix 2 with $0 \le f_{\zeta}
\le 1$, being the relative contribution of technileptons into the
total dark matter density, and  $S_2$ is the mass of $\zeta$ in
units of 100 GeV). By definition, $f_{\zeta}=1-x$, where $x$ was
defined in Eq.~(\ref{x_definition}).
The $\bar \zeta^{++}$ expels the $^4He$
from the $(^4He^{++}\zeta^{--})$ and annihilates in
$\zeta$-positronium $(\bar \zeta^{++} \zeta^{--})$.

The process of $\bar{\zeta}^{++}$ capture by the
$(^4He^{++}\zeta^{--})$ atom looks as follows \cite{FKS}. Being in
thermal equilibrium with the plasma, the free $\bar{\zeta}^{++}$
have momentum $k = \sqrt{2T m_{\zeta}}$. If their wavelength is
much smaller than the size of the $(He^{++}\zeta^{--})$ atom, they
can penetrate inside the atom and bind with $A^{--}$, expelling
the $^4He$ from it. The rate of this process is determined by the
size of the $(He^{++}\zeta^{--})$ atoms and is given by \beq \sv_0
\sim \pi R_{o} ^2 \sim \frac{\pi}{(\bar \alpha m_{He})^2} =
\frac{\pi}{2 I_{o} m_{He}} \approx 3 \cdot
10^{-15}\frac{\text{cm}^3}{\text{s}}. \label{capzeta} \eeq Here
$\bar \alpha = Z_{\zeta} Z_{He} \alpha.$ At temperature $T< T_a =
\bar \alpha^2 m_{He} \frac{m_{He}}{2m_{\zeta}} =
\frac{I_{o}m_{He}}{m_{\zeta}} = 4 \cdot 10^{-2} I_{o}/S_2$, the
wavelength $\lambda=1/k$ of $\bar{\zeta}^{++}$, exceeds the size
$R_o=1/(\bar \alpha m_{\zeta})$ of the $(He^{++}\zeta^{--})$
``atom". The rate of the $(He^{++}\zeta^{--})$ catalysis is
suppressed by a factor $(R_{o}/\lambda)^3 = (T/T_a)^{3/2}$ and is
given by \beq \sv_{cat}(T<T_a)= \sv_0 (T/T_a)^{3/2}.
\label{capzeta1} \eeq The
decrease of
the antiparticle abundance $r_{\bar{\zeta}}$ is described by
 \beq \frac{dr_{\bar{\zeta}}}{dx}=f_{1He}\left<\sigma
v\right> r_{\bar{\zeta}}(r_{\bar{\zeta}} + \kappa),
\label{AEsrecomb} \eeq where $x=T/I_{o}$, $r_{He\zeta}=r_{\zeta}$,
$\sv$ is given by Eqs.~(\ref{capzeta}) at $T>T_a$ and
(\ref{capzeta1}) at $T<T_a$. The solution of this equation is
given in \cite{FKS} and has the form \beq
r_{\bar{\zeta}}=\frac{\kappa\cdot r_{f\bar{\zeta}}}{(\kappa
+r_{f\bar{\zeta}}) \exp\left( \kappa J_{o} \right) -
r_{f\bar{\zeta}}}, \label{rHeEbar} \eeq where $r_{f\bar{\zeta}}$
is
the frozen concentration of $\bar{\zeta}$ and \beq J_{o} =
\int_0^{x_{fHe}} f_{1He}\left<\sigma v\right>dx
 = m_{Pl} (\frac{\pi}{2m_{He}}) \cdot x_{fHe}\approx 1.4 \cdot
10^{17}, \label{Jo} \eeq  where $x_{fHe}=1/27.$ The factor in the
exponent is $\kappa J_{o} = 6 \cdot 10^{5} f_{\zeta}/S_2 $. It
leads to a huge exponential suppression of the antiparticles at
all reasonable values of $\kappa$ and $S_2$.

\subsection{\label{EHESBBN} Techni-O-helium in the  SBBN}
The formation of techni-O-helium reserves a fraction of $^4He$ and
thus it changes the primordial abundance of $^4He$. For the
lightest possible masses of the techniparticles $m_{\zeta} \sim
m_{TB} \sim 100 \GeV$, this effect can reach 50\% of the $^4He$
abundance formed in SBBN. Even if the mass of the techniparticles
is of the order of TeV, $5\%$ of the $^4He$ abundance is hidden in
the techni-O-helium atoms. This can lead to important consequences
once we compare the SBBN theoretical predictions to observations.

The question of the participation of techni-O-helium in nuclear
transformations and its direct influence on the chemical element
production is less evident. Indeed, techni-O-helium looks like an
$\alpha$ particle with a shielded electric charge. It can closely
approach nuclei due to the absence of a Coulomb barrier. Because
of this, it seems that in the presence of techni-O-helium, the
character of SBBN processes should change drastically. However, it
might not be the case.

The following simple argument can be used to indicate that the
techni-O-helium influence on SBBN transformations might not lead
to binding of $A^{--}$ with nuclei heavier than $^4He$. In fact,
the size of techni-O-helium is of the order of the size of $^4He$
and for a nucleus $^A_ZQ$ with electric charge $Z>2$, the size of
the Bohr orbit for an $Q A^{--}$ ion is less than the size of the
nucleus $^A_ZQ$. This means that while binding with a heavy
nucleus, $A^{--}$ penetrates it and interacts effectively with a
part of the nucleus of a size less than the corresponding Bohr
orbit. This size corresponds to the size of $^4He$, making
techni-O-helium the most bound $Q A^{--}$ atomic state. It favors
a picture, according to which a techni-O-helium collision with a
nucleus, results in the formation of techni-O-helium and the whole
process looks like an elastic collision.

The interaction of the $^4He$ component of $(He^{++}A^{--})$ with
a $^A_ZQ$ nucleus can lead to a nuclear transformation due to the
reaction \beq ^A_ZQ+(HeA) \rightarrow ^{A+4}_{Z+2}Q +A^{--},
\label{EHeAZ} \eeq provided that the masses of the initial and
final nuclei satisfy the energy condition \beq M(A,Z) + M(4,2) -
I_{o}> M(A+4,Z+2), \label{MEHeAZ} \eeq where $I_{o} = 1.6 \MeV$ is
the binding energy of techni-O-helium and $M(4,2)$ is the mass of
the $^4He$ nucleus.

This condition is not valid for stable nuclei participating in
reactions of the SBBN. However, tritium $^3H$, which is also
formed in SBBN with abundance $^3H/H \sim 10^{-7}$ satisfies this
condition and can react with techni-O-helium, forming $^7Li$ and
opening the path of successive techni-O-helium catalyzed
transformations to heavy nuclei. This effect might strongly
influence the chemical evolution of matter on the pre-galactic
stage and needs a self-consistent consideration within the Big
Bang nucleosynthesis network. However, the following arguments
show that this effect may not lead to immediate contradiction with
observations as it might be expected.
\begin{itemize}
\item[$\bullet$] On the path of reactions (\ref{EHeAZ}), the final
nucleus can be formed in the excited $(\alpha, M(A,Z))$ state,
which can rapidly experience an $\alpha$- decay, giving rise to
techni-O-helium regeneration and to an effective quasi-elastic
process of $(^4He^{++}A^{--})$-nucleus scattering. It leads to a
possible suppression of the techni-O-helium catalysis of nuclear
transformations. \item[$\bullet$] The path of reactions
(\ref{EHeAZ}) does not stop on $^7Li$ but goes further through
$^{11}B$, $^{15}N$, $^{19}F$, ... along the table of the chemical
elements. \item[$\bullet$] The cross section of reactions
(\ref{EHeAZ}) grows with the mass of the nucleus, making the
formation of the heavier elements more probable and moving the
main output away from a potentially dangerous Li and B
overproduction.
\end{itemize}

The first publications on possible realistic composite dark matter
scenarios \cite{I,FKS} gave rise to the development of another
aspect of the problem, the Charged massive particles BBN (CBBN),
studying the influence of unstable negatively charged massive
particles in BBN
\cite{Pospelov:2006sc,Kaplinghat,Kohri:2006cn,Kino,Kawasaki:2007xb,Jedamzik:2007cp}.
The important difference of CBBN considered in these papers, from
our approach, is that singly charged particles $X^-$ with charge
$-1$ do not screen the $+2$ charge of $He$ in a $(HeX)^+$ ion-like
bound system, and the Coulomb barrier of the $(HeX)^+$ ion can
strongly hamper the path for the creation of isotopes, heavier
than $^6Li$. Therefore, $^6Li$ created in the $D+(HeX)$ reaction
cannot dominantly transform into heavier elements and if not
destructed by $X$-decay products, it should remain in the
primordial chemical content. It makes the $^6Li$ overproduction
found in \cite{Pospelov:2006sc} a really serious trouble for a
wide range of parameters for unstable $X$ particles.

It should be noted that the approach of \cite{Pospelov:2006sc} is
not supported by \cite{Kohri:2006cn}. Moreover, we can mention the
following effects \cite{FKS}, missed in its solution for the $^7Li$ problem:
(i) the competitive process of $^7Li$ creation by a similar
mechanism in the reaction $^3H+(HeX)^+$  with tritium and (ii) the
effects of non-equilibrium nucleosynthesis reactions, induced by
hadronic and electromagnetic cascades from products of $X$ decays.
The latter effect, which was discussed in \cite{Kohri:2006cn},
implies a self-consistent treatment based on the theory of
non-equilibrium cosmological nucleosynthesis
\cite{Linde3,Linde4,bookKh} (see also
\cite{Kawasaki:2004yh,Kawasaki:2004qu,Kohri:2005wn,Jedamzik:2006xz}).
Both effects (i) and (ii) were not studied in
\cite{Pospelov:2006sc}.

The amount of techni-O-helium in our scenario formally exceeds by
a few orders of magnitude the constraint $n_X/s \le 10^{-17}$,
derived for concentration $n_X$ of metastable $X^-$ particles in
the units of entropy density $s$ in Eq.~(10) of
\cite{Pospelov:2006sc}. However, it should be noted that this
constraint is not valid for our case if the binding energy $I_o=
1589 \keV$ of techni-O-helium is taken into account. According to
\cite{Kohri:2006cn}, this approximation is valid for $0<ZZ_{TC}
\alpha M_Z R_Z<1$, where $R_Z \sim 1.2 A^{1/3}/200 \MeV^{-1}$ is
the size of nucleus, which is the case for the $(HeA)$ atom. Then
the $D+(HeA) -> ^6Li + A$ reaction, which the constraint is based
on, does not occur. This reaction can take place only if the
account for charge distribution in the $He$ nucleus
\cite{Pospelov:2006sc} reduces the binding energy of $(HeA)$ down
to $E= 1200 \keV$ or $E= 1150 \keV$ as discussed in \cite{FKS}.
Then this channel becomes possible, but similar to the case of
tritium, the chain of techni-O-helium transformations
(\ref{EHeAZ}), started from deuterium does not stop on $^6Li$, but
goes further through $^{10}B$, $^{14}N$, $^{18}F$, ... along the
table of the chemical elements. Such a qualitative change of the
physical picture appeals to necessity in a detailed nuclear
physics treatment of the ($A^{--}$+ nucleus) systems and of the
whole set of transformations induced by techni-O-helium, including
an analysis of possible fast conversion of helium to carbon and of
the formation  of a $(^8BeA^{--})$ system, discussed in \cite{FKS}
as potential dangers for our approach. Though the above arguments
do not seem to make these dangers immediate and obvious, a
detailed study of this complicated problem is needed.

\section{\label{Interactions} Techni-O-helium Universe}
\subsection{Gravitational instability of the techni-O-helium gas}
 Due to nuclear interactions of its helium constituent with
nuclei in cosmic plasma, the techni-O-helium gas is in thermal
equilibrium with plasma and radiation on the Radiation Dominance
(RD) stage, and the energy and momentum transfer from the plasma
is effective. The radiation pressure acting on plasma is then
effectively transferred to density fluctuations of techni-O-helium
gas and transforms them in acoustic waves at scales up to the size
of the horizon. However, as it was first noticed in \cite{I}, this
transfer to heavy nuclear-interacting species becomes ineffective
before the end of the RD stage and such species decouple from
plasma and radiation. Consequently, nothing prevents the
development of gravitational instability in the gas of these
species. This argument is completely applicable to the case of
techni-O-helium.

At temperature $T < T_{od} \approx 45 S^{2/3}_2\eV$, first
estimated in \cite{I} for the case of OLe-helium, the energy and
momentum transfer from baryons to techni-O-helium is not effective
because $n_B \sv (m_p/m_o) t < 1$, where $m_o$ is the mass of the
$tOHe$ atom and $S_2=\frac{m_o}{100 \GeV}$. Here \beq \sigma
\approx \sigma_{o} \sim \pi R_{o}^2 \approx
10^{-25}\cm^2\label{sigOHe}, \eeq and $v = \sqrt{2T/m_p}$ is the
baryon thermal velocity. The techni-O-helium gas decouples from
the plasma and plays the role of dark matter, which starts to
dominate in the Universe at $T_{RM}=1 \eV$.

The development of gravitational instabilities of the
techni-O-helium gas triggers large scale structure formation, and
the composite nature of techni-O-helium makes it more close to
warm dark matter.

The total mass of the $tOHe$ gas with density $\rho_d =
\frac{T_{RM}}{T_{od}} \rho_{tot}$ within the cosmological horizon
$l_h=t$ is
$$M=\frac{4 \pi}{3} \rho_d t^3.$$ In the period of decoupling $T = T_{od}$, this mass  depends
strongly on the techniparticle mass $S_2$ and is given by \beq
M_{od} = \frac{T_{RM}}{T_{od}} m_{Pl} (\frac{m_{Pl}}{T_{od}})^2
\approx 2 \cdot 10^{46} S^{-8/3}_2 \g = 10^{13} S^{-8/3}_2
M_{\odot}, \label{MEPm} \eeq where $M_{\odot}$ is the solar mass.
The techni-O-helium is formed only at $T_{rHe}$ and its total mass
within the cosmological horizon in the period of its creation is
$M_{o}=M_{od}(T_{o}/T_{od})^3 = 10^{37} \g$.

On the RD stage before decoupling, the Jeans length $\lambda_J$ of
the $tOHe$ gas was of the order of the cosmological horizon
$\lambda_J \sim l_h \sim t.$ After decoupling at $T = T_{od}$, it
falls down to $\lambda_J \sim v_o t,$ where $v_o =
\sqrt{2T_{od}/m_o}.$ Though after decoupling the Jeans mass in the
$tOHe$ gas correspondingly falls down
$$M_J \sim v_o^3 M_{od}\sim 3 \cdot 10^{-14}M_{od},$$ one should
expect strong suppression of fluctuations on scales $M<M_o$, as
well as adiabatic damping of sound waves in the RD plasma for
scales $M_o<M<M_{od}$. It provides suppression of small scale
structure in the considered model for all reasonable masses of
techniparticles.

The cross section of mutual collisions of techni-O-helium
``atoms'' is given by Eq.~(\ref{sigOHe}). The $tOHe$ ``atoms" can
be considered as collision-less gas in clouds with a number
density $n_{o}$ and a size $R$, if $n_{o}R < 1/\sigma_{o}$. This
condition is valid for the techni-O-helium gas in galaxies.

Mutual collisions of techni-O-helium ``atoms'' determine the
evolution timescale for a gravitationally bound system of
collision-less $tOHe$ gas $$t_{ev} = 1/(n \sigma_{o} v) \approx 2
\cdot 10^{20} (1 \cm^{-3}/n)^{7/6}\s,$$ where the relative
velocity $v = \sqrt{G M/R}$ is taken for a cloud of mass $M_o$ and
an internal number density $n$. This timescale exceeds
substantially the age of the Universe and the internal evolution
of techni-O-helium clouds cannot lead to the formation of dense
objects. Being decoupled from baryonic matter, the $tOHe$ gas does
not follow the formation of baryonic astrophysical objects (stars,
planets, molecular clouds...) and forms dark matter halos of
galaxies.

\subsection{Techniparticle component of cosmic rays}
 The nuclear interaction of techni-O-helium with cosmic rays gives rise to
ionization of this bound state in the interstellar gas and to
acceleration of free techniparticles in the Galaxy. During the
lifetime of the Galaxy $t_G \approx 3 \cdot 10^{17} \s$, the
integral flux of cosmic rays $$F(E>E_0)\approx 1 \cdot
\left(\frac{E_0}{1 \GeV} \right)^{-1.7} \cm^{-2} \s^{-1}$$ can
disrupt the fraction of galactic techni-O-helium $ \sim
F(E>E_{min}) \sigma_o t_G \le 10^{-3},$ where we took $E_{min}
\sim I_o.$ Assuming a universal mechanism of cosmic ray
acceleration, a universal form of their spectrum, taking into
account that the $^4He$ component corresponds to $\sim 5$\% of the
proton spectrum, and that the spectrum is usually reduced to the
energy per nucleon, the anomalous low $Z/A$ $-2$ charged
techniparticle component can be present in cosmic rays at a level
of \beq
 \frac{A^{--}}{He} \ge 3 \cdot 10^{-7} \cdot S_2^{-3.7}. \eeq
 This flux may be within the reach for PAMELA and AMS02 cosmic ray
experiments.

Recombination of free techniparticles with protons and nuclei in
the interstellar space can give rise to radiation in the range
from few tens of keV - 1 MeV. However such a radiation is below
the cosmic nonthermal electromagnetic background radiation
observed in this range.

\subsection{\label{EpMeffects} Effects of techni-O-helium catalyzed processes in the Earth}
The first evident consequence of the proposed excess is the
inevitable presence of $tOHe$ in terrestrial matter. This is
because terrestrial matter appears opaque to $tOHe$ and stores all
its in-falling flux.

If the $tOHe$ capture by nuclei is not effective, its diffusion in
matter is determined by elastic collisions, which have a transport
cross section per nucleon \beq \sigma_{tr} = \pi R_{o}^2
\frac{m_{p}}{m_o} \approx 10^{-27}/S_2 \cm^{2}. \label{sigpEpcap}
\eeq In atmosphere, with effective height $L_{atm} =10^6 \cm$ and
baryon number density $n_B= 6 \cdot 10^{20} \cm^{-3}$, the opacity
condition $n_B \sigma_{tr} L_{atm} = 6 \cdot 10^{-1}/S_2$ is not
strong enough. Therefore, the in-falling $tOHe$ particles are
effectively slowed down only after they fall down terrestrial
surface in $16 S_2$ meters of water (or $4 S_2$ meters of rock).
Then they drift with velocity $V = \frac{g}{n \sigma v} \approx 8
S_2 A^{1/2} \cm/\s$ (where $A \sim 30$ is the average atomic
weight in terrestrial surface matter, and $g=980~
\text{cm}/\text{s}^2$), sinking down the center of the Earth on a
timescale $t = R_E/V \approx 1.5 \cdot 10^7 S_2^{-1}\s$, where
$R_E$ is the radius of the Earth.

The in-falling techni-O-helium flux from dark matter halo is
$\mathcal{F}=n_o v_h/8 \pi$, where the number density of $tOHe$ in
the vicinity of the Solar System is $n_o=3 \cdot 10^{-3}S_2^{-1}
\cm^{-3}$ and the averaged velocity $v_h \approx 3 \cdot
10^{7}\cm/\s$. During the lifetime of the Earth ($t_E \approx
10^{17} \s$), about $2 \cdot 10^{38}S_2^{-1}$ techni-O-helium
atoms were captured. If $tOHe$ dominantly sinks down the Earth, it
should be concentrated near the Earth's center within a radius
$R_{oc} \sim \sqrt{3T_c/(m_o 4 \pi G \rho_c)}$, which is $\le 10^8
S_2^{-1/2}\cm$, for the Earth's central temperature $T_c \sim 10^4
\K$ and density $\rho_c \sim 4 \g/\cm^{3}$.

Near the Earth's surface, the techni-O-helium abundance is
determined by the equilibrium between the in-falling and
down-drifting fluxes. It gives $$n_{oE}=2 \pi \mathcal{F}/V = 3
\cdot 10^3 \cdot S_2^{-2}\cdot A^{-1/2}\cm^{-3},$$ or for $A \sim
30$ about $5 \cdot 10^2 \cdot S_2^{-2} \cm^{-3}$. This number
density corresponds to the fraction $$f_{oE} \sim 5 \cdot 10^{-21}
\cdot S_2^{-2}$$ relative to the number density of the terrestrial
atoms $n_A \approx 10^{23} \cm^{-3}$.

These neutral $(^4He^{++}A^{--})$ ``atoms" may provide a catalysis
of cold nuclear reactions in ordinary matter (much more
effectively than muon catalysis). This effect needs a special and
thorough investigation. On the other hand, if $A^{--}$ capture by
nuclei, heavier than helium, is not effective and does not lead to
a copious production of anomalous isotopes, the
$(^4He^{++}A^{--})$ diffusion in matter is determined by the
elastic collision cross section (\ref{sigpEpcap}) and may
effectively hide techni-O-helium from observations.

One can give the following argument for an effective regeneration
and quasi-elastic collisions of techni-O-helium in terrestrial
matter. The techni-O-helium can be destroyed in the reactions
(\ref{EHeAZ}). Then, free $A^{--}$ are released and due to a
hybrid Auger effect (capture of $A^{--}$, ejection of ordinary $e$
from the atom with atomic number $A$, and charge of the nucleus
$Z$), $A^{--}$-atoms are formed, in which $A^{--}$ occupies highly
an excited level of the $(_Z^AQA)$ system, which is still much
deeper than the lowest electronic shell of the considered atom.
The $(_Z^AQA)$ atomic transitions to lower-lying states cause
radiation in the intermediate range between atomic and nuclear
transitions. In course of this falling down to the center of the
$(Z-A^{--})$ system, the nucleus approaches $A^{--}$. For $A>3$
the energy of the lowest state $n$ (given by $E_n=\frac{M \bar
\alpha^2}{2 n^2} = \frac{2 A m_p Z^2 \alpha^2}{n^2}$)  of the
$(Z-A^{--})$ system (having reduced mass $M \approx A m_p$) with a
Bohr orbit $r_n=\frac{n}{M \bar \alpha}= \frac{n}{2 A Z m_p
\alpha}$, exceeding the size of the nucleus $r_A \sim
A^{1/3}m^{-1}_{\pi}$ ($m_{\pi}$ being the mass of the pion), is
less than the binding energy of $tOHe$. Therefore the regeneration
of techni-O-helium in a reaction, inverse to (\ref{EHeAZ}), takes
place. An additional reason for the domination of the elastic
channel of the reactions (\ref{EHeAZ}) is that the final state
nucleus is created in the excited state and its de-excitation via
$\alpha$-decay can also result in techni-O-helium regeneration. If
regeneration is not effective and $A^{--}$ remains bound to the
heavy nucleus, anomalous isotope of $Z-2$ element should appear.
This is a serious problem for the considered model.

However, if the general picture of sinking down is valid, it might
give no more than the ratio $f_{oE} \sim 5 \cdot 10^{-21} \cdot S_2^{-2}$
of number density of anomalous isotopes to the number density of atoms of
terrestrial matter around us, which is below the experimental
upper limits for elements with $Z \ge 2$. For comparison, the best
upper limits on the anomalous helium were obtained in \cite{exp3}.
It was found, by searching with the use of laser spectroscopy for
a heavy helium isotope in the Earth's atmosphere, that in the mass
range 5 GeV - 10000 GeV, the terrestrial abundance (the ratio of
anomalous helium number to the total number of atoms in the Earth)
of anomalous helium is less than $2 \cdot 10^{-19}$ - $3 \cdot
10^{-19}$.

\subsection{\label{DMdirect} Direct search for techni-O-helium}

It should be noted that the nuclear cross section of the
techni-O-helium interaction with matter escapes the severe
constraints \cite{McGuire:2001qj} on strongly interacting dark
matter particles (SIMPs) \cite{Starkman,McGuire:2001qj} imposed by
the XQC experiment \cite{XQC}.

In underground detectors, $tOHe$ ``atoms'' are slowed down to
thermal energies and give rise to energy transfer $\sim 2.5 \cdot
10^{-3} \eV A/S_2$, far below the threshold for direct dark matter
detection. It makes this form of dark matter insensitive to the
CDMS constraints. However, $tOHe$ induced nuclear transformation
can result in observable effects.

 Therefore, a special strategy of such a search is needed, that
can exploit sensitive dark matter detectors
 on the ground or in space. In particular, as it was revealed in
\cite{Belotsky:2006fa}, a few $\g$ of superfluid $^3He$ detector
\cite{Winkelmann:2005un}, situated in ground-based laboratory can
be used to put constraints on the in-falling techni-O-helium flux
from the galactic halo.
\section{\label{Discussion} Discussion}
In this paper we explored the cosmological implications of a
 walking technicolor model with doubly charged technibaryons and
technileptons. The considered model escapes most of the drastic
problems of the Sinister Universe \cite{Glashow}, related to the
primordial $^4He$ cage for $-1$ charge particles and a consequent
overproduction of anomalous hydrogen \cite{Fargion:2005xz}. These
charged $^4He$ cages pose a serious problem for composite dark
matter models with single charged particles, since their Coulomb
barrier prevents successful recombination of positively and
negatively charged particles. The doubly charged $A^{--}$
techniparticles considered in this paper, bind with $^4He$ in the
 techni-O-helium neutral states.

To avoid overproduction of anomalous isotopes, an excess of $-2$
charged techniparticles over their antiparticles should be
generated in the Universe. In all the previous realizations of
composite dark matter scenarios, this excess was put by hand to
saturate the observed dark matter density. In our paradigm, this
abundance of techibaryons and/or technileptons is connected
naturally to the baryon relic density.

A challenging problem that we leave for future work is the nuclear
transformations, catalyzed by techni-O-helium. The question about
their consistency with observations remains open, since special
nuclear physics analysis is needed to reveal what are the actual
techni-O-helium effects in SBBN and in terrestrial matter. Another
aspect of the considered approach is more clear. For reasonable
values of the techiparticle mass, the amount of primordial $^4He$,
bound in this atom like state is significant and should be taken
into account in comparison to observations.

The destruction of techni-O-helium by cosmic rays in the Galaxy
releases free charged techniparticles, which can be accelerated
and contribute to the flux of cosmic rays. In this context, the
search for techniparticles at accelerators and in cosmic rays
acquires the meaning of a crucial test for the existence of the
basic components of the composite dark matter. At accelerators,
techniparticles would look like stable doubly charged heavy
leptons, while in cosmic rays, they represent a heavy $-2$ charge
component with anomalously low ratio of electric charge to mass.

To conclude, walking technicolor cosmology can naturally resolve
most of problems of composite dark matter. Therefore, the model
considered in this paper with stable $-2$ charged particles might
provide a realistic physical basis for a composite dark matter
scenario.

\section*{Acknowledgements}

The work of C.K. is supported by the Marie Curie Fellowship under
contract MEIF-CT-2006-039211. We are grateful to K.M. Belotsky and
F.Sannino for reading the manuscript and for important remarks.
M.Kh. is grateful to CERN (Geneve, Switzerland) and to ICTP
(Trieste, Italy) for hospitality.
\section*{ Appendix 1. Charge asymmetry in freezing out of particles and antiparticles
} The frozen number density of cosmic relics, which were in
equilibrium with the primordial plasma, is conventionally deduced
\footnote{We follow here the results obtained in
\cite{Fargion:2005xz,FKS} with the help of K.M. Belotsky} from the
equation \cite{ZeldNov}
\begin{equation}
\dot{n}+3Hn=\left<\sigma_{ann}v\right>(n_{eq}^2-n^2). \label{sym}
\end{equation}
This equation is written for the case of a charge symmetry of the
particles in question, i.e. for the case when number densities of
particles $X$ and antiparticles $\bar{X}$ are equal
$n_X=n_{\bar{X}}=n$. The value $n_{eq}$ corresponds to their
equilibrium number density and is given by the Boltzmann
distribution
\begin{equation}
n_{eq}=g_S \left(\frac{mT}{2\pi}\right)^{3/2}\exp \left(-\frac{m}{T}\right).
\label{neq}
\end{equation}
Here $g_S$ and $m$ are the number of spin states and the mass of the
given particle.

During the cooling, $n_{eq}$ decreases exponentially and becomes,
below the freezing out temperature $T_f$, much smaller than the
real density $n$, so the term
$\left<\sigma_{ann}v\right>n_{eq}^2$, describing the creation of
$X\bar{X}$ from the plasma can be neglected \cite{Turner}. It
allows to obtain an approximate solution of Eq.~(\ref{sym}).

In case of a charge asymmetry one needs to split Eq.~(\ref{sym})
in two: for $n_X$ and $n_{\bar{X}}$, which are not equal now.
\begin{eqnarray}
\dot{n}_X+3Hn_X=\left<\sigma_{ann}v\right>(n_{eq\,X}n_{eq\,\bar{X}}-n_{X}n_{\bar{X}}),\nonumber\\
\dot{n}_{\bar{X}}+3Hn_{\bar{X}}=\left<\sigma_{ann}v\right>(n_{eq\,X}n_{eq\,\bar{X}}-n_{X}n_{\bar{X}}).
\label{asym}
\end{eqnarray}
The values $n_{eq\,X}$ and $n_{eq\,\bar{X}}$ are given by
Eq.~(\ref{neq}) with inclusion of the chemical potential, which
for $X$ and for $\bar{X}$ are related as
$\mu_X=-\mu_{\bar{X}}=\mu$ (see, e.g., \cite{Dolgov}). So
\begin{equation}
n_{eq\,X,\bar{X}}=\exp\left(\pm\frac{\mu}{T}\right) n_{eq},
\label{nmueq}
\end{equation}
where upper and lower signs are for $X$ and $\bar{X}$ respectively.
So
\begin{equation}
n_{eq\,X}n_{eq\,\bar{X}}= n_{eq}^2. \label{neq2}
\end{equation}
A degree of asymmetry will be described in the conventional manner
(as for baryons) by the ratio of the difference between $n_{X}$ and
$n_{\bar{X}}$ to the number density of relic photons at the modern
period
\begin{equation}
\kappa_{\gamma\,mod}=\frac{n_{X\,mod}-n_{\bar{X}\,mod}}{n_{\gamma\,mod}}.
\label{kappagamma}
\end{equation}
However, for practical purposes it is more suitable to use the
ratio to the entropy density, which unlike Eq.~(\ref{kappagamma}),
does not change in time provided entropy conservation. The photon
number density $n_{\gamma}$ and the entropy density $s$ are given
by
\begin{equation}
n_{\gamma}=\frac{2\zeta(3)}{\pi^2}T^3,\;\;\; s=\frac{2\pi^2
g_s}{45}T^3=1.80g_sn_{\gamma}, \label{ngammas}
\end{equation}
where
\begin{equation}
g_s=\sum_{bos}
g_S(\frac{T_{bos}}{T})^3+\frac{7}{8}\sum_{ferm}g_S(\frac{T_{ferm}}{T})^3.
\label{gs}
\end{equation}
The sums in Eq.~(\ref{gs}) are over ultrarelativistic bosons and
fermions. So
\begin{equation}
\kappa=\frac{n_{X}-n_{\bar{X}}}{s},\;\;\;
\kappa=\frac{\kappa_{\gamma\,mod}}{1.8g_{s\,mod}}, \label{kappa}
\end{equation}
where $g_{s\,mod}=43/11 \approx 3.91$.
Eq.~(\ref{kappa}) provides a connection between $n_{X}$ and
$n_{\bar{X}}$. Let us pass to the variables
\begin{equation}
r_+=\frac{n_X}{s},\;\;\; r_-=\frac{n_{\bar{X}}}{s},\;\;\;
r=\frac{n_X+n_{\bar{X}}}{s},\;\;\; x=\frac{T}{m}. \label{rx}
\end{equation}
The apparent relations between the $r_i$ are
\begin{equation}
r_+-r_-=\kappa,\;\;\; r_++r_-=r. \label{r-r-r}
\end{equation}
Provided that the essential entropy redistribution does not take
place ($g_s=\text{const.}$) during the period of freezing out, a
transformation to the variable $x$ is possible $$-Hdt=dT/T=dx/x.$$
On the RD stage the Hubble parameter depends on $T$ as
\begin{equation}
H=\frac{2\pi}{3} \sqrt{\frac{\pi g_{\epsilon}}{5}}
\frac{T^2}{m_{Pl}}, \label{Heps}
\end{equation}
where $g_{\epsilon}$ is given by
\begin{equation}
g_{\epsilon}=\sum_{bos}
g_S(\frac{T_{bos}}{T})^4+\frac{7}{8}\sum_{ferm}g_S(\frac{T_{ferm}}{T})^4.
\label{geps}
\end{equation}
For $r_+$, $r_-$ and $r$ from Eqs.~(\ref{asym}) one obtains the
equations
\begin{eqnarray}
\frac{dr_+}{dx}=f_1\left<\sigma_{ann}v\right>\left( r_+(r_+-\kappa)-f_2(x)\right)\nonumber\\
\frac{dr_-}{dx}=f_1\left<\sigma_{ann}v\right>\left( r_-(r_-+\kappa)-f_2(x)\right)\nonumber\\
\frac{dr}{dx}=\frac{1}{2}f_1\left<\sigma_{ann}v\right>\left(
r^2-\kappa^2-4f_2(x)\right). \label{drrr}
\end{eqnarray}
Here
\begin{eqnarray}
f_1&=&\frac{s}{Hx},\nonumber\\
f_2(x)&=&\frac{n_{eq}^2}{s^2}=\frac{45^2 g_S^2}{2^5\pi^7g_s^2
x^3}\exp\left(-\frac{2}{x}\right). \label{f12}
\end{eqnarray}
By using Eqs.~(\ref{ngammas}) and Eq.~(\ref{Heps}), one finds that
on the RD stage $f_1$ is
$$f_1=\sqrt{\frac{\pi g_s^2}{45g_{\epsilon}}}m_{Pl}m$$
and independent of $x$.

To solve Eqs.~(\ref{drrr}) analogously to Eq.~(\ref{sym}), namely
neglecting $f_2(x)$ in them, starting with some $x=x_f$, it would
not be more difficult  to define the moment $x=x_f$.
Nonetheless, if one supposes that such a moment is defined, $r_i$
will be
\begin{eqnarray}
r_+(x\approx 0)=\frac{\kappa\cdot r_{+f}}{r_{+f}-(r_{+f}-\kappa) \exp\left(-\kappa J\right)},\nonumber\\
r_-(x\approx 0)=\frac{\kappa\cdot r_{-f}}{(\kappa+r_{-f}) \exp\left( \kappa J \right)-r_{-f}},\\
r(x\approx 0)=\kappa \frac{(\kappa+r_{f})\exp\left( \kappa J
\right)+r_f-\kappa} {(\kappa+r_{f})\exp\left( \kappa J
\right)-(r_f-\kappa)}.\nonumber \label{rrr}
\end{eqnarray}
Here $r_{i\,f}=r_i(x=x_f)$,
$$J=\int_0^{x_f} f_1\left<\sigma_{ann}v\right>dx.$$
All $r_i$ (at any moment) are related with the help of
Eqs.~(\ref{r-r-r}). Taking into account Eq.~(\ref{nmueq}) or
Eq.~(\ref{neq2}) for $r_{i\,f}$ one obtains
\begin{eqnarray}
r_{\pm\,f}&=&\frac{1}{2}\left(\sqrt{4f_2(x_f)+\kappa^2}\pm \kappa
\right),
\nonumber \\
r_{f}&=&\sqrt{4f_2(x_f)+\kappa^2}. \label{rpm}
\end{eqnarray}
For $\left<\sigma_{ann}v\right>$ independent of $x$ on the RD
stage, $f_1$ also independent of $x$, and $x_f$ defined from the
condition $R(T_f)=H(T_f)$ for the reaction rate
$R(T_f)=n_{eq}(T_f)\left<\sigma_{ann}v(T_f)\right>$, leading to
$$n_{eq}(T_f)\left<\sigma_{ann}v(T_f)\right>/H(T_f)=\frac{n_{eq}}{s}\cdot \frac{s}{H x_f}\cdot \left<\sigma_{ann}v(x_f)\right> \cdot x_f =$$
\beq = \sqrt{f_2(x_f)} f_1 \left<\sigma_{ann}v(x_f)\right> \cdot x_f
=1, \label{f2f1H} \eeq one obtains \beq \sqrt{f_2(x_f)} =
\frac{1}{f_1 \left<\sigma_{ann}v\right> \cdot x_f} = \frac{1}{J}.
\label{f2f1J} \eeq
If (a) $\left<\sigma_{ann}v\right>=\alpha^2/m^2$ or (b)
$\left<\sigma_{ann}v\right>=\alpha/\sqrt{Tm^3}$ and one assumes
$f_1=\text{const}$, then
\begin{eqnarray}
J_a=\sqrt{\frac{\pi g_s^2}{45g_{\epsilon}}}m_{Pl}\frac{\alpha^2}{m}x_f, \nonumber\\
J_b=\sqrt{\frac{\pi
g_s^2}{45g_{\epsilon}}}m_{Pl}\frac{\alpha}{m}2\sqrt{x_f}.
\end{eqnarray}
\section*{Appendix 2. Primordial technileptons from the Big Bang Universe}
As already mentioned, the minimal walking technicolor model
considered in this paper can allow the creation of $\zeta$ excess
that might contribute (or even saturate) to the modern dark matter
density in the form of techni-O-helium ``atoms". For light baryon
excess $\eta_B= n_{B\,mod}/n_{\gamma \,mod} = 6 \cdot 10^{-10}$,
it gives a $\zeta$-excess
 \begin{eqnarray}
 \eta_{\zeta}=n_{\zeta\,mod}/n_{\gamma \,mod}
 = 3 \cdot 10^{-11}f_{\zeta}
(\frac{100{\GeV}}{m_{\zeta}}), \label{excess}
\end{eqnarray}
where $m_{\zeta}$ is the mass of $\zeta$. By definition
$f_{\zeta}$ is the contribution of $\zeta$ into the modern dark
matter density $f_{\zeta}=1-x$, where $x$ was defined in
Eq.~(\ref{x_definition}). For future use, following
\cite{Glashow,Fargion:2005xz}, it is convenient to relate the
baryon density $\Omega_B=0.044$ and the technilepton density
$\Omega_{L'}=0.224$ with the entropy density $s$, and to introduce
$r_B = n_B/s$ and $r_{\zeta}=n_{\zeta}/s$. Taking into account
that $s_{mod}=7.04\cdot n_{\gamma\,mod},$ one obtains $r_B \sim 8
\cdot 10^{-11}$ and \beq r_{\zeta} =4 \cdot 10^{-12}
\left(\frac{100{\GeV}}{m_{\zeta}}\right)= 4 \cdot 10^{-12}
f_{\zeta} /S_2. \label{sexcess} \eeq

\subsection*{\label{Chronology} Chronological cornerstones of the technilepton-Universe}
After the generation of technilepton asymmetry, the thermal
history of technileptons in chronological order looks as follows:

1) $10^{-10}S_2^{-2}\s \le t \le 6 \cdot10^{-8}S_2^{-2}\s$ at
$m_{\zeta} \ge T \ge T_f=m_{\zeta}/31 \approx 3 S_2 \GeV.$
$\zeta$-lepton pair $\zeta \bar \zeta$ annihilation and freezing out.
For large $m_{\zeta}$
the abundance of frozen out $\zeta$-lepton pairs is not suppressed
in spite of an $\zeta$-lepton excess.

2)$ t \sim 2.4 \cdot 10^{-3}S_2^{-2}\s$  at $T \sim I_{\zeta} = 20
S_2 \MeV.$ The temperature corresponds to the binding energy
$I_{\zeta} = Z_{\zeta}^4 \alpha^2 m_{\zeta}/4 \approx 20 S_2 \MeV$
($Z_{\zeta}=2$) of $\zeta$-positronium ``atoms" $(\zeta^{--} \bar
\zeta^{++})$, in which $\bar \zeta^{++}$ annihilate. At large
$m_{\zeta}$ this annihilation is not at all effective to reduce
the $\zeta \bar \zeta$ pairs abundance. These pairs are eliminated
in the course of the successive evolution of $\zeta$-matter.

3)$100\s \le t \le 300\s$  at $100 \keV\ge T \ge I_{o}/27 \approx
60 \keV.$ $^4He$ is formed in the SBBN and virtually all free
$\zeta^{--}$ are trapped by $^4He$ in $(^4He^{++}\zeta^{--})$.
Note that in the period $100 \keV \le T \le 1.6 \MeV$, $^4He$ is
not formed, therefore it is only after {\it the first three
minutes}, when $(^4He^{++}\zeta^{--})$ trapping of $\zeta^{--}$
can take place. Being formed, techni-OLe-helium catalyzes the
binding of free $\bar \zeta^{++}$ with its constituent
$\zeta^{--}$ into $\zeta$-positronium and complete annihilation of
all the primordial antiparticles. At large $m_{\zeta}$, effects of
$(\zeta^{--} \bar \zeta^{++})$ annihilation, catalyzed by
techni-O-helium, do not cause any contradictions with
observations.

4) $t \sim 10^{12}\s$  at $T \sim T_{RM} \approx 1 \eV.$ The
 techniparticle dominance starts with techni-O-helium ``atoms"
playing the role more close to warm dark matter in the formation
of large scale structures.

\subsection*{\label{Efreezing} Freezing out of $\bold{\zeta}$-leptons}

In the early Universe at temperatures highly above their masses,
the $\zeta$-fermions  were in thermodynamical equilibrium with the
relativistic plasma. It means that at $T>m_{\zeta}$ the excessive
$\zeta$ were accompanied by $\zeta \bar \zeta$ pairs.

During the expansion, when the temperature $T$ falls
 below the mass of the $\zeta$-particles, the concentration of particles and
antiparticles is given by the equilibrium. The equilibrium
concentration of $\zeta \bar \zeta$ pairs starts to decrease at
$T<m_{\zeta}=100 S_2 \GeV$. At the freezing out temperature $T_f$,
the rate of expansion exceeds the rate of annihilation to photons
$\zeta \bar \zeta  \rightarrow \gamma \gamma$, to $W,Z$-bosons
$\zeta \bar \zeta \rightarrow WW (ZZ)$, or to pairs of light
fermions $f$ (quarks and charged leptons) $\zeta \bar \zeta
\rightarrow \bar f f$ (the latter takes place both due to
electromagnetic and weak interactions). Then $\zeta$ leptons and
their antiparticles $\bar \zeta$ are frozen out.

In the case of freezing out of $\zeta$-leptons one has (see Appendix
1)
$$f_{1\zeta}=\sqrt{\frac{\pi g_s^2}{45g_{\epsilon}}}m_{Pl}m_{\zeta} \approx 2.5 m_{Pl}m_{\zeta},$$
$\left<\sigma_{ann}v\right> =\frac{\bar \alpha^2}{m_{\zeta}^2}$
and \beq J_{\zeta}=\sqrt{\frac{\pi
g_s^2}{45g_{\epsilon}}}m_{Pl}\frac{\bar \alpha^2}{m_{\zeta}}x_f,
\label{JU} \eeq where $\bar \alpha=Z_{\zeta}^2\alpha + \bar
\alpha_{ew}$, $Z_{\zeta}=2$ is the absolute value of the electric
charge of the $\zeta$ and $\bar \alpha_{ew}$ takes into account
the effects of $W$ and $Z$ bosons in technilepton annihilation. By
putting in Eq.~(\ref{f12}) $g_S=2$, $g_s \sim 100$, one obtains
the solution of the transcendent equation (\ref{f2f1J})
$$x_f \approx \left(\ln{\left(\frac{45 g_S}{2^{5/2}\pi^{7/2}g_s} \cdot f_{1\zeta} \left<\sigma_{ann}v\right>\right)}\right)^{-1} \approx$$
$$ \approx \frac{1}{30}\cdot \frac{1}{(1 - \ln{(S_2)}/30)}.$$
Taking $g_s \approx g_{\epsilon} \sim 100$, one finds from
Eq.~(\ref{JU}) $J_{\zeta} = 6.5 \cdot 10^{13}/S_2 (1 -
\ln{(S_6)}/30)^{-1}$ and from Eq.~(\ref{f2f1J}) $\sqrt{4f_2(x_f)}
= 2/J_{\zeta} =3 \cdot 10^{-13} S_2\cdot (1 - \ln{(S_2)}/30)$. For
$\kappa = r_{\zeta}= 4 \cdot 10^{-12} f_{\zeta}/S_2$, one has
$\kappa J_{\zeta} = 26 f_{\zeta}/S_2^2$. At $S_2<2.7$ $4f_2(x_f) <
\kappa^2$ and $r_{\pm\,f}$ is given by Eq.~(\ref{rpm}). Since
$4f_2(x_f) \gg \kappa^2$ for $S_2 \gg 1$, one obtains from
Eq.~(\ref{rpm}) \beq
r_{\pm\,f}=\frac{1}{2}\left(\sqrt{4f_2(x_f)}\pm \kappa \right).
\label{rpmf} \eeq The frozen out abundances of $\zeta$-leptons and
their antiparticles are given by
\begin{eqnarray}
r_{\zeta}=\frac{\kappa\cdot r_{+f}}{r_{+f}-(r_{+f}-\kappa) \exp\left(-\kappa J_{\zeta}\right)}=F_{\zeta}(S_2),\nonumber\\
r_{\bar{\zeta}}=\frac{\kappa\cdot r_{-f}}{(\kappa+r_{-f}) \exp\left(
\kappa J_{\zeta} \right)-r_{-f}} = F_{\bar{\zeta}}(S_2).
\label{rUpm}
\end{eqnarray}
For growing $S_2 \gg 1$, the solution Eq.~(\ref{rUpm}) approaches
the values
\begin{eqnarray}
r_{\zeta} \approx \sqrt{f_2(x_f)} + \kappa/2 \approx\nonumber\\
\approx 1.5 \cdot 10^{-13}S_2\cdot (1 - \ln{(S_2)}/30) + 2 \cdot 10^{-12} f_{\zeta}/S_2,\nonumber\\
r_{\bar{\zeta}} \approx \sqrt{f_2(x_f)} - \kappa/2 \approx \nonumber\\
\approx 1.5 \cdot 10^{-13}S_2\cdot (1 - \ln{(S_2)}/30) - 2 \cdot
10^{-12} f_{\zeta}/S_2. \label{rUSpm}
\end{eqnarray}
At $S_2 < 5 f_{\zeta}$, the factor in the exponent $\kappa
J_{\zeta}$ exceeds 1, and some suppression of the
$(\bar{\zeta})$-abundance takes place. For $S_2$ close to 1, one
has
\begin{eqnarray}
r_{\zeta}=F_{\zeta}(S_2) \approx \kappa = 4 \cdot 10^{-12}f_{\zeta}/S_2,\nonumber\\
r_{\bar{\zeta}}=F_{\bar{\zeta}}(S_2) \approx 5 \cdot 10^{-3}\kappa
S_2^4 \exp\left( -26 f_{\zeta}/S_2^2 \right). \label{rU1pm}
\end{eqnarray}
At $S_2=1$, the factor in the exponent reaches the value $\kappa
J_{\zeta}=26 f_{\zeta}$ and the solution Eq.~(\ref{rUpm}) gives
$r_{\zeta} \approx \kappa = 4 \cdot 10^{-12} f_{\zeta}$ and
\begin{eqnarray}
r_{\bar{\zeta}} &\approx& \frac{\kappa\cdot
r_{-f}}{\kappa+r_{-f}}\exp\left(- \kappa J_{\zeta} \right) \approx
r_{-f} \exp\left(- \kappa J_{\zeta} \right)
\nonumber \\
&\approx& 10^{-14} \exp\left( -26 f_{\zeta} \right) \approx 3 \cdot
10^{-25} \nonumber
\end{eqnarray}
for $f_{\zeta}=1$ and $r_{-f}
\approx 10^{-14}$ from Eq.~(\ref{rpm}).

The $S_2$-dependence of the frozen out abundances (in units of the
entropy density) of the $\zeta$ leptons and their antiparticles
are
\begin{eqnarray}
r_{\zeta}= F_{\zeta}(S_2),\nonumber\\
r_{\bar{\zeta}}=F_{\bar{\zeta}}(S_2), \label{rEfr}
\end{eqnarray}
given by Eq.~(\ref{rUpm}). For growing $S_2 \gg 1$, the solution
Eq.~(\ref{rUpm}) approaches the values
\begin{eqnarray}
r_{AC} \approx \sqrt{f_2(x_f)} + \kappa/2 \approx\nonumber\\
\approx 1.5 \cdot 10^{-13}S_2\cdot (1 - \ln{(S_2)}/30) + 2 \cdot 10^{-12}f_{\zeta}/S_2,\nonumber\\
r_{\bar{AC}} \approx \sqrt{f_2(x_f)} - \kappa/2 \approx \nonumber\\
\approx 1.5 \cdot 10^{-13}S_2\cdot (1 - \ln{(S_2)}/30) - 2 \cdot
10^{-12} f_{\zeta}/S_2. \label{rESfr}
\end{eqnarray}
At $S_2 < 5 f_{\zeta}$, there is a exponential suppression of the
$\bar{\zeta}$ abundance. For $S_2$ close to 1, one has
\begin{eqnarray}
r_{\zeta}=F_{\zeta}(S_2) \approx \kappa = 4 \cdot 10^{-12}/S_2,\nonumber\\
r_{\bar{\zeta}}=f_{\bar{\zeta}}(S_2) \approx 5 \cdot 10^{-3}\kappa
S_2^4 \exp\left( -26 f_{\zeta}/S_2^2 \right). \label{rE1fr}
\end{eqnarray}
At $S_2=1$, the solution Eq.~(\ref{rE1fr}) gives
$$r_{\zeta} \approx \kappa_A = \kappa_C = 2 \cdot 10^{-12}f_{\zeta},$$ and
$$r_{\bar{\zeta}} \approx 3 \cdot 10^{-25}.$$
On the other hand at $S_2 > 5$, the concentration of frozen out
$\zeta$-lepton pairs exceeds the one of the $\zeta$-lepton excess
given by Eq.~(\ref{sexcess}) and this effect grows with $S_2$ as
$\propto S_2^2$ at large $S_2$. So in this moment, in spite of an
assumed $\zeta$-lepton asymmetry, the frozen out concentration of
antiparticles $\bar \zeta$ is not strongly suppressed and they
cannot be neglected in the cosmological evolution of
$\zeta$-matter.

The antiparticles $\bar \zeta$ should be effectively annihilated
in the successive processes of $\zeta \bar \zeta$ recombination in
bound $(\zeta \bar \zeta)$ $\zeta$-positronium states.

\subsection*{\label{anE} $\bar \zeta$ annihilation in $\zeta$-positronium states}
The frozen out antiparticles $\bar \zeta$ can bind at $T<
I_{\zeta}$ with the corresponding particles $\zeta$ into
positronium-like systems and annihilate. The binding is provided
by the Coulomb interaction of electromagnetic charges
$Z_{\zeta}=2$. Since the lifetime of these positronium-like
systems is much less than the timescale of their disruption by
energetic photons, the direct channel of $\bar \zeta$ and $\bar
\zeta$ binding in $(\zeta \bar \zeta)$, followed by a rapid
annihilation, cannot be compensated by an inverse reaction of
photo-ionization. That is why $\bar \zeta$ begins to bind with
$\zeta$ and annihilates as soon as the temperature becomes less
than $I_{\zeta}$. The decrease of the $\bar \zeta$ abundance due
to the $\zeta \bar \zeta$ recombination is governed by the
equation \beq \frac{dr_{\bar \zeta}}{dt} = -r_{\zeta} r_{\bar
\zeta} \cdot s \cdot \sv, \label{hadrecombE} \eeq where $s$ is the
entropy density and
$$\sv=  (\frac{16\pi}{3^{3/2}}) \cdot \frac{\bar \alpha}{T^{1/2}\cdot m_{\zeta}^{3/2}}.$$
Here $\bar \alpha = Z_{\zeta}^2 \alpha$.

In the analysis of various recombination processes, we can use the
interpolation formula for the recombination cross section deduced
in \cite{Q,Fargion:2005xz,FKS}: \beq
 \sigma_r=(\frac{2\pi}{3^{3/2}}) \cdot \frac{\bar \alpha ^3}{T\cdot I_1} \cdot \log{(\frac{I_1}{T})}\eeq
and the recombination rate given by \cite{Q,Fargion:2005xz}\beq
 \sv=(\frac{2\pi}{3^{5/2}}) \cdot \frac{\bar \alpha ^3}{T\cdot I_1} \cdot \log{(\frac{I_1}{T})} \cdot
 \frac{k_{in}}{M}.
\label{recdisc} \eeq Here $k_{in}= \sqrt{2 T M}$, $I_1 \approx
\bar \alpha^2 M/2$ is the ionization potential, and $M$ is the
reduced mass for a pair of recombining particles. The constant
$\bar \alpha$ for recombining particles with charges $Z_1$ and
$Z_2$ is related to the fine structure constant $\alpha$ via $\bar
\alpha= Z_1 Z_2 \alpha$. The approximation Eq.~(\ref{recdisc})
followed from the known result for electron-proton recombination
\beq \sigma_{rec}=\sigma_r
  =\sum_i \frac{8\pi}{3^{3/2}} \bar \alpha^3 \frac{e^4}{Mv^2i^3} \frac{1}{(Mv^2/2+I_i)},
\label{recep} \eeq where $v$ is the velocity of the particles.
 $I_i$ is the ionization potential  ($I_i=I_1/i^2$). The index $i$
 runs from one to infinity.

 To sum approximately over $i$, it was noted in  \cite{Q,FKS}, that $\sigma_r\propto 1/i$
 for $I_i >> Mv^2/2=T_{eff}$, while at $I_i<T_{eff}$, the cross section
 $\sigma_i\propto 1/i^3$ falls down rapidly.

Using the formalism of Appendix 1, we can rewrite
Eq.~(\ref{hadrecombE}) as \beq \frac{dr_{\bar \zeta}}{dx}=f_{1
\bar \zeta}\left<\sigma v\right> r_{\bar \zeta}(r_{\bar
\zeta}+\kappa), \label{Esrecomb1} \eeq where $x=T/I_{\zeta}$, the
asymmetry $\kappa = r_{\zeta}-r_{\bar \zeta} = 4 \cdot
10^{-12}f_{\zeta} /S_2$ is given by Eq.~(\ref{sexcess}) and
$$f_{1 \bar \zeta}=\sqrt{\frac{\pi g_s^2}{45g_{\epsilon}}}m_{Pl}I_{\zeta} \approx m_{Pl}I_{\zeta}.$$
The concentration of the remaining $\bar \zeta$ is given by
Eq.~(\ref{rrr}) of Appendix 1 \beq r_{\bar
\zeta}=\frac{\kappa\cdot r_{f\bar \zeta}}{(\kappa+r_{f\bar \zeta})
\exp\left( \kappa J_{\zeta} \right) - r_{f\bar \zeta}},
\label{rEbar} \eeq where $r_{f \bar \zeta}$ is given by
Eq.~(\ref{rEfr}) and
$$J_{\zeta}=\int_0^{x_{f\bar \zeta}} f_{1\bar \zeta}\left<\sigma v\right>dx =$$
\beq = m_{Pl}I_{\zeta} 4 \pi (\frac{2}{3^{3/2}}) \cdot \frac{ \bar
\alpha^2}{I_{\zeta}\cdot m_{\zeta}} \cdot 2 \cdot x_{f\bar
\zeta}^{1/2}\approx 0.8 \cdot 10^{15}/S_2. \label{JEbar} \eeq In
the evaluation of Eq.~(\ref{JEbar}) we took into account that the
decrease of $\bar \zeta$ starts at $T\sim I_{\zeta}$, so that
$x_{f\bar \zeta} \sim 1$. At $S_2 < 57f_{\zeta}^{1/2}$, the
abundance of $\bar \zeta$ is suppressed exponentially.

Indeed, one has $\kappa J_{\zeta} \approx 3200 f_{\zeta}/S_2^2$ in
the exponent of Eq.~(\ref{rEbar}). Similar to the case of the
AC-leptons \cite{FKS}, it differs significantly from the situation
revealed in \cite{Fargion:2005xz} for the tera-positrons in
Glashow's sinister model \cite{Glashow}. Though in both cases a
decrease of antiparticles due to the formation of positronium like
systems is induced by electromagnetic interaction and the factor
in the exponent is determined by the square of the fine structure
constant $\alpha$, in the case of $\zeta$-leptons, this factor is
enhanced by $Z_{\zeta}^4=16$ times due to the $Z_{\zeta}^4$
dependence of $\bar \alpha^2$. It results in a much wider mass
interval for $\zeta$-leptons, in which the primordial pair
abundance is exponentially suppressed.

At $S_2$ close to 1,  the condition $r_{f\bar \zeta} \ll \kappa$
in the solution Eq.~(\ref{rEbar}) provides the approximate
solution
$$r_{\bar \zeta}= r_{f\bar \zeta} \cdot  \exp\left(- \kappa J_{\zeta} \right) \approx
10^{-14}S_2^3 \exp\left( -3200 f_{\zeta}/S_2^2 \right).$$
For $S_2>5$, the condition $r_{f\bar \zeta} \gg \kappa$ is valid.
Therefore the solution Eq.~(\ref{rEbar}) has the form \beq r_{\bar
\zeta} \approx \frac{\kappa}{ \exp\left( \kappa J_{\zeta} \right)
- 1}, \label{rEbars} \eeq which gives for $S_2 <
57f_{\zeta}^{1/2}$
$$r_{\bar \zeta}= \kappa \cdot  \exp\left(- \kappa J_{\zeta} \right)
\approx \left(\frac{2\cdot 10^{-12}}{S_2}\right) \exp\left(
-3200f_{\zeta}/S_2^2\right).$$ At large $S_2 > 57f_{\zeta}^{1/2}$,
the approximate solution is given by
$$r_{\bar \zeta} \approx \frac{1}{ J_{\zeta} }  -  \frac{\kappa}{ 2 }
\approx 1.25 \cdot 10^{-15} S_2 - 2 \cdot 10^{-12}f_{\zeta}/S_2.$$
In the result, the residual amount of $\bar \zeta$ remains at $S_2
> 57f_{\zeta}^{1/2}$ enormously high, being  for  $S_2 > 70f_{\zeta}^{1/2}$
larger than the AC-lepton excess. This effect grows with $S_2 >
70f_{\zeta}^{1/2}$ as $\propto S_2^2.$

The general expression for the $\zeta$ abundance $r_{\zeta}$ after
the $\zeta$-positronium annihilation has the form (see
Eq.~(\ref{rrr}) of Appendix 1)
$$r_{\zeta}=\frac{\kappa\cdot r_{\zeta f}}{r_{\zeta f}-(r_{\zeta f}-\kappa) \exp\left(-\kappa J_{\zeta}\right)},$$
where $J_{\zeta}$ is given by Eq.~(\ref{JEbar}) and $r_{\zeta f}$
is given by  Eq.~(\ref{rEfr}).
With the account for $r_{\zeta f} > \kappa$, for all $S_2$, one
obtains \beq r_{\zeta}=\frac{\kappa}{1-\exp\left(-\kappa
J_{\zeta}\right)}. \label{rEtpan} \eeq This gives $r_{\zeta}
\approx 1/J_{\zeta}  + \kappa/2 \approx 3 \cdot 10^{-16} S_2 + 2
\cdot 10^{-12}f_{\zeta}/S_2$ for large $S_2$, and  $\kappa$ for
$S_2 < 57f_{\zeta}^{1/2}$.

\end{document}